\documentclass[twocolumn]{aastex62}

\usepackage{graphicx}

\usepackage[utf8]{inputenc}
\usepackage{subfigure}
\usepackage[T1]{fontenc}

\begin{document}
\title{The Panchromatic Hubble Andromeda Treasury: Triangulum Extended Region (PHATTER) II. The Spatially Resolved Recent Star Formation History of M33}
\author[0000-0003-3252-352X]{Margaret Lazzarini}
\affiliation{California Institute of Technology, 1200 E California Blvd., Pasadena, CA 91125, USA}
\affiliation{Department of Astronomy, Box 351580, University of Washington, Seattle, WA 98195, USA}

\author[0000-0002-7502-0597]{Benjamin F. Williams}
\affiliation{Department of Astronomy, Box 351580, University of Washington, Seattle, WA 98195, USA}

\author[0000-0001-7531-9815]{Meredith J. Durbin}
\affiliation{Department of Astronomy, Box 351580, University of Washington, Seattle, WA 98195, USA}

\author[0000-0002-1264-2006]{Julianne J.\ Dalcanton}
\affiliation{Department of Astronomy, Box 351580, University of Washington, Seattle, WA 98195, USA}
\affiliation{Center for Computational Astrophysics, Flatiron Institute, 162 Fifth Avenue, New York, NY 10010, USA}

\author[0000-0003-2599-7524]{Adam Smercina}
\affiliation{Department of Astronomy, Box 351580, University of Washington, Seattle, WA 98195, USA}

\author[0000-0002-5564-9873]{Eric F.\ Bell}
\affiliation{Department of Astronomy, University of Michigan, 323 West Hall, 1085 S. University Ave., Ann Arbor, MI 48105, USA}

\author[0000-0003-1680-1884]{Yumi Choi}
\affiliation{Space Telescope Science Institute, 3700 San Martin Dr., Baltimore, MD 21218, USA}

\author[0000-0001-8416-4093]{Andrew Dolphin}
\affiliation{Raytheon, Tucson, AZ 85726, USA}
\affiliation{Steward Observatory, University of Arizona, Tucson, AZ 85726, USA}

\author[0000-0003-0394-8377]{Karoline Gilbert}
\affiliation{Space Telescope Science Institute, 3700 San Martin Dr., Baltimore, MD 21218, USA}
\affiliation{The William H. Miller III Department of Physics \& Astronomy, Bloomberg Center for Physics and Astronomy, Johns Hopkins University, 3400 N. Charles Street, Baltimore, MD 21218}

\author[0000-0001-8867-4234]{Puragra Guhathakurta}
\affiliation{UCO/Lick Observatory, Department of Astronomy \& Astrophysics, University of California Santa Cruz, 1156 High Street, Santa Cruz,
California 95064, USA}

\author[0000-0002-5204-2259]{Erik Rosolowsky}
\affiliation{Department of Physics, University of Alberta, Edmonton, AB T6G 2E1, Canada}

\author[0000-0003-0605-8732]{Evan Skillman}
\affiliation{School of Physics and Astronomy, University of Minnesota, 116 Church St. SE, Minneapolis, MN 55455, USA}
\affiliation{Minnesota Institute for Astrophysics, University of Minnesota, 116 Church St. SE, Minneapolis, MN 55455, USA}

\author[0000-0003-4122-7749]{O. Grace Telford}
\affiliation{Rutgers University, Department of Physics and Astronomy, 136 Frelinghuysen Road, Piscataway, NJ 08854, USA}

\author[0000-0002-6442-6030]{Daniel Weisz}
\affiliation{Department off Astronomy, University of California Berkeley, Berkeley, CA 94720, USA}


\correspondingauthor{Margaret Lazzarini}
\email{mlazz@caltech.edu}

\begin{abstract}
We measure the spatially resolved recent star formation history (SFH) of M33 using optical images taken with the Hubble Space Telescope as part of the Panchromatic Hubble Andromeda Treasury: Triangulum Extended Region (PHATTER) survey. The area covered by the observations used in this analysis covers a de-projected area of $\sim$38 kpc$^{2}$ and extends to $\sim$3.5 and $\sim$2 kpc from the center of M33 along the major and semi-major axes, respectively. We divide the PHATTER optical survey into 2005 regions that measure 24 arcsec, $\sim$100 pc, on a side and fit color magnitude diagrams for each region individually to measure the spatially resolved SFH of M33 within the PHATTER footprint. There are significant fluctuations in the SFH on small spatial scales and also galaxy-wide scales that we measure back to about 630 Myr ago. We observe a more flocculent spiral structure in stellar populations younger than about 80 Myr, while the structure of the older stellar populations is dominated by two spiral arms. We also observe a bar in the center of M33, which dominates at ages older than about 80 Myr. Finally, we find that the mean star formation rate (SFR) over the last 100 Myr within the PHATTER footprint is 0.32$\pm$0.02 M$_{\odot}$ yr$^{-1}$. We measure a current SFR (over the last 10 Myr) of 0.20$\pm$0.03 M$_{\odot}$ yr$^{-1}$. This SFR is slightly higher than previous measurements from broadband estimates, when scaled to account for the fraction of the D25 area covered by the PHATTER survey footprint.
\end{abstract}


\section{Introduction}
The recent star formation history (SFH) of a galaxy allows us to look back in time to see the effects of the astrophysical phenomena that have influenced its evolution. If this recent SFH can also be spatially resolved, we can see the impact of the astrophysics that controls the stellar structure of the galaxy including the formation and evolution of spiral arms and the connection between local star formation events and the surrounding dust and gas from which new stars form. 

There are several methods for deriving the recent SFH of a galaxy. For studies of distant galaxies, integrated light from unresolved stars can relate a galaxy's current star formation rate to other galaxy-scale properties such as its mass, luminosity, and color. While these studies are useful for obtaining statistically significant results with large samples of galaxies, they do not allow us to study the physics surrounding star formation within each galaxy in detail. For this, we must turn to the more local universe where we can resolve star formation across individual galaxies rather than approximating their physical properties with a handful of parameters. 

Most methods for measuring the recent SFH of local galaxies rely on star formation rate (SFR) indicators, which trace the presence of young, newly formed stars through a variety of signatures. These tracers can be both direct and indirect, each with a characteristic time scale. Ultraviolet (UV) emission from young massive stars traces star formation within the last 100$-$200 Myr, corresponding to the lifetimes of these stars \citep{Kennicutt&Evans2012}. As a direct tracer of star formation, UV emission has been widely used to measure star formation rates in nearby galaxies, especially after the launch of GALEX \citep[e.g.,][]{GildePaz2007}. However, UV flux is attenuated by interstellar dust, making it more of a lower limit on the SFR \citep[e.g.,][]{Bell2003}. H$\alpha$ emission from gas ionized by young O-stars traces very recent star formation within the last $\sim$5 Myr \citep{Byler2017}. H$\alpha$ is thus an indirect tracer of young, massive stars ($>$15 M$_{\odot}$), but is one of the best available measure of the instantaneous SFR in external galaxies. Dust attenuation \citep{Kennicutt&Evans2012} and the leakage of ionizing photons \citep{Lee2009,Choi2020} are sources of systematic uncertainty in H$\alpha$ measurements. 

A more direct method for measuring the star formation rate is to not just confirm the presence of young, massive stars, but instead to actually count the number of stars of various masses directly and use this distribution to infer time-resolved star formation rates that would produce the observed population \citep[e.g.,][]{Dolphin2002,Weisz2008,Williams2011,Kennicutt&Evans2012}. This method requires deep, high resolution observations to resolve individual stars for a full population, but provides the information necessary to produce a spatially-resolved star formation history. We can use resolved stars to construct color-magnitude diagrams (CMDs) for specific spatial regions within galaxies. The distribution of stars on the CMD is a product of the galaxy's history of star formation, the evolution of the galaxy's metallicity, and the initial mass function (IMF), all observed through the distribution of dust. Like all SFH measurement techniques, modelling the CMD relies on assumptions about the IMF, stellar evolution and atmospheric models, binary fraction and dust. While CMD fitting is is not the only method we currently have to produce time-resolved SFR measurements -- SED-fitting can also produce time-resolved SFR measurements -- it does produce the highest time-resolution in these measurements.

Much of the early work using CMD fitting on galaxies in the local universe focused on low mass galaxies and smaller portions of large galaxies. Dwarf galaxies are an excellent target for resolved stellar populations work because of their abundance in the Local Volume, making them close enough to resolve their individual stars \citep[e.g.,][]{Zaritsky&Harris2004,Harris&Zaritsky2009,McQuinn2010,Weisz2011,Weisz2014,Geha2015,Cigoni2019}. Studies focused on larger galaxies have often relied on small fields across the disk to answer questions about galaxy evolution, such as inside-out growth. However, these small fields make resolving the temporal evolution of large scale structures such as spiral arms or bars difficult \citep[e.g.,][]{Williams2002,Williams2003,Brown2006,Brown2007,Brown2008,Williams2009,Williams2010,Bernard2012,Bernard2015,Choi2015}. 

CMD modelling has successfully derived spatially resolved SFHs for well-resolved galaxies with wider field coverage with multiple HST fields. There have been several significant applications of this SFH measurement technique in nearby galaxies including M31 \citep{Lewis,Williams2017}, M33 \citep{Williams2009}, M81 \citep{Choi2015}, NGC 300 \citep{Gogarten2010}, and NGC 2403 \citep{Williams2013NGC2403}. The common distance of stars in these nearby galaxies allows for a more simplified CMD modelling approach over stellar populations in the Milky Way.

\citet{Lewis} provided the first spatially resolved star formation history of a significant portion of an L$_{*}$ galaxy, M31, using photometry from the Panchromatic Hubble Andromeda Treasury (PHAT). Their fits were optimized for main sequence stars, providing a recent star formation history for roughly 9000 regions measuring $\sim$100 pc by 100 pc across a 0.5 square degree area in the northern third of the star forming disk within the last $\sim$500 Myr. \citet{Williams2017} provided an analogous measurement of the spatially resolved ancient star formation history of M31 by focusing their fits on older stellar populations. Both studies capitalized on the rich dataset from PHAT, which provided six-band photometry from the near-infrared through near-UV for over 100 million individual stars in M31 \citep{Dalcanton,WilliamsPHAT}.

The spatially resolved SFH maps of M31 derived via CMD-modelling have had several significant applications in subsequent work. The \citet{Lewis} maps were important in establishing that M31 had a disk wide burst of star formation when it merged \citep{DSouza&Bell2018,Hammer2018}. By comparing the spatially resolved SFH maps derived via CMD-modelling with SFRs derived from various indicators, \citet{Lewis2017} found that the SFR indicators in M31 were under-estimating the total amount of star formation. \citet{DiazRodriguez2018} used SFH maps to measure the progenitor mass distribution for core collapse supernovae in M31 and M33 using the SFH maps of \citet{Lewis} and \citet{Jennings2014}. The \citet{Lewis} SFH maps in M31 have also been used to measure the age distribution of high mass X-ray binaries in M31 \citep{Williams2018,Lazzarini2018,Lazzarini2021}.

The insight into M31's evolution gained from the PHAT survey has been incredibly valuable. In this paper, we derive the spatially resolved recent SFH of M33. Though the techniques are the same, M33 and M31 are very different galaxies (e.g., in mass, structure, and star-forming activity), providing important context to the results from PHAT. The star formation rate intensity of M33 is higher than M31 \citep{Verley2009, Lewis}, allowing us to study high intensity star formation regions. M33 also has well known stellar and gas-phase metallicity gradients \citep{Cioni2009,Magrini2009,Magrini2010,beasley2015,ToribioSanCipriano2016,Lin2017}, which are similar to the LMC \citep{carrera2008}. Unlike M31, M33 has likely not undergone a recent significant merger event, though recent spectroscopic observations may hint at a centrally-concentrated population of accreted stars \citep{gilbert2021}. Its warped outer disk does suggest a mild encounter with M31 in the last 1--3 Gyr \citep{putman2009}, but proper motions measurements indicate that M33 is likely on its fits infall into the M31 system \citep[e.g.,][]{Patel2017a,Patel2017b,vanderMarel2019}. This makes it an excellent target for CMD-fitting based SFH measurements, allowing results to be compared directly to the SFHs of disrupting Magellanic spirals such as the LMC \citep{mazzi2021}, and undisturbed field spirals such as NGC 300 \citep{Gogarten2010}.

M33 is of comparable mass and metallicity to the Large Magellanic Cloud (LMC; $M_{\star}\,{\sim}\,3{\times}10^9\,M_{\odot}$, [M/H]\,$\sim$\,$-$0.5; e.g., \citealt{vandermarel2002,carrera2008,vandermarel2012,beasley2015}), making it one of the most massive satellite galaxies in the Local Group. It does not have a significant bulge component \citep{Kormendy&McClure1993,McLean&Liu1996} and has been observed to host a weak bar \citep{Elmegreen1992,Regan&Vogel1994,Block2004,Hernandez-Lopez2009,Corbelli2007}, both of which reduce confusion between bulge and disk stellar populations. Its proximity allows us to resolve stars down to the ancient main sequence \citep{Williams2009}. In fact, we can resolve its stellar populations better than we can in M31 due to its similar distance but lower stellar surface density. 

There has been significant previous work to measure the spatially-resolved SFR of M33 using star formation indicators across the electromagnetic spectrum and resolved stellar populations, and we briefly review them here. These include the spatially resolved SFR of M33 using broadband flux SF indicators and resolved stellar populations CMD-modelling.

The spatially resolved SFR of M33 has been previously studied using star formation indicators including FUV, H$\alpha$, and 24 $\mu$m emission. \citet{Verley2009} measured the global SFR of M33 over the last 100 Myr using extinction corrected FUV and H$\alpha$ flux and detected a radial decline in SFR. \citet{Verley2007} measured the global and local SFR of M33 using 24 $\mu$m emission from HII regions. Far-infrared emission from dust is an important SF indicator because much of the FUV flux from young stars is absorbed and re-emitted by surrounding dust \citep{Boquien2010, Boquien2011}. There can be large discrepancies between SF measurements with FUV, H$\alpha$, and far-infrared SF indicators \citep{Boquien2015}. This highlights the importance of CMD modelling, which does not rely on the calibration of SF indicators and can also produce time-resolved measurements. However, calibrating SF indicators with nearby galaxies such as M33 is very important work because we can only measure SFHs with CMD modelling for galaxies that are near enough to resolve their stellar populations.

CMD-fitting has previously been used to measure the recent SFH of M33 in targeted regions of the disk, confirming a mean age and metallicity gradient in the disk, suggesting inside-out growth \citep{Barker2007,Williams2009, Barker2011}. CMD-based SFH measurements of SFHs around supernova remnants in M33 and M31 suggest that some massive stars may not explode as supernovae \citep{Jennings2014}. These CMD-based SFH measurements were targeted and did not cover the full disk, or a large contiguous area of the disk of M33. The PHATTER data now allow such a measurement for the first time, opening the door to studies of temporal evolution in morphological features. 

In this work we generate maps of SFH within the last $\sim$630 Myr in M33 using optical photometry from the Panchromatic Hubble Andromeda Treasury: Triangulum Extended Region (PHATTER) survey \citep{Williams2021}. Our approach mirrors that in the \citet{Lewis} analysis with the PHAT data in M31, allowing the two measurements to be directly compared. This work is one of the first projects using the rich PHATTER dataset to study the star formation, dust, morphology, and more with photometry for over 22 million individual stars in M33.

In this paper we use optical photometry from the PHATTER survey \citep{Williams2021} to derive the spatially resolved recent star formation history of M33. In Section \ref{sec:data} we present the optical photometry used as well as our method for creating artificial stars to measure photometric errors. In Section \ref{sec:derivation_sfhs} we present our method for deriving the spatially resolved recent SFH of M33. In Section \ref{sec:results} we present the resulting SFH maps and in Section \ref{sec:discussion} we discuss our results. In Section \ref{sec:conclusions} we summarize our findings.

For the analysis presented in this paper, we assume a distance modulus for M33 of 24.67 \citep{deGrijs2014}, which is equivalent to a distance of $\sim$859 kpc. We use the measured D25 ellipse presented by \citet{GildePaz2007} which is centered at $1^{h}33^{m}50^{s}.9$, $30^{\circ}39^{\prime}35^{\prime \prime}.8$. We adopt a position angle (P.A.) of $201^{\circ}$ and an inclination angle of $55^{\circ}$ \citep{Koch2018}. 

\section{PHATTER Data}\label{sec:data}
\begin{figure*}
\centering
\includegraphics[width=0.95\textwidth]{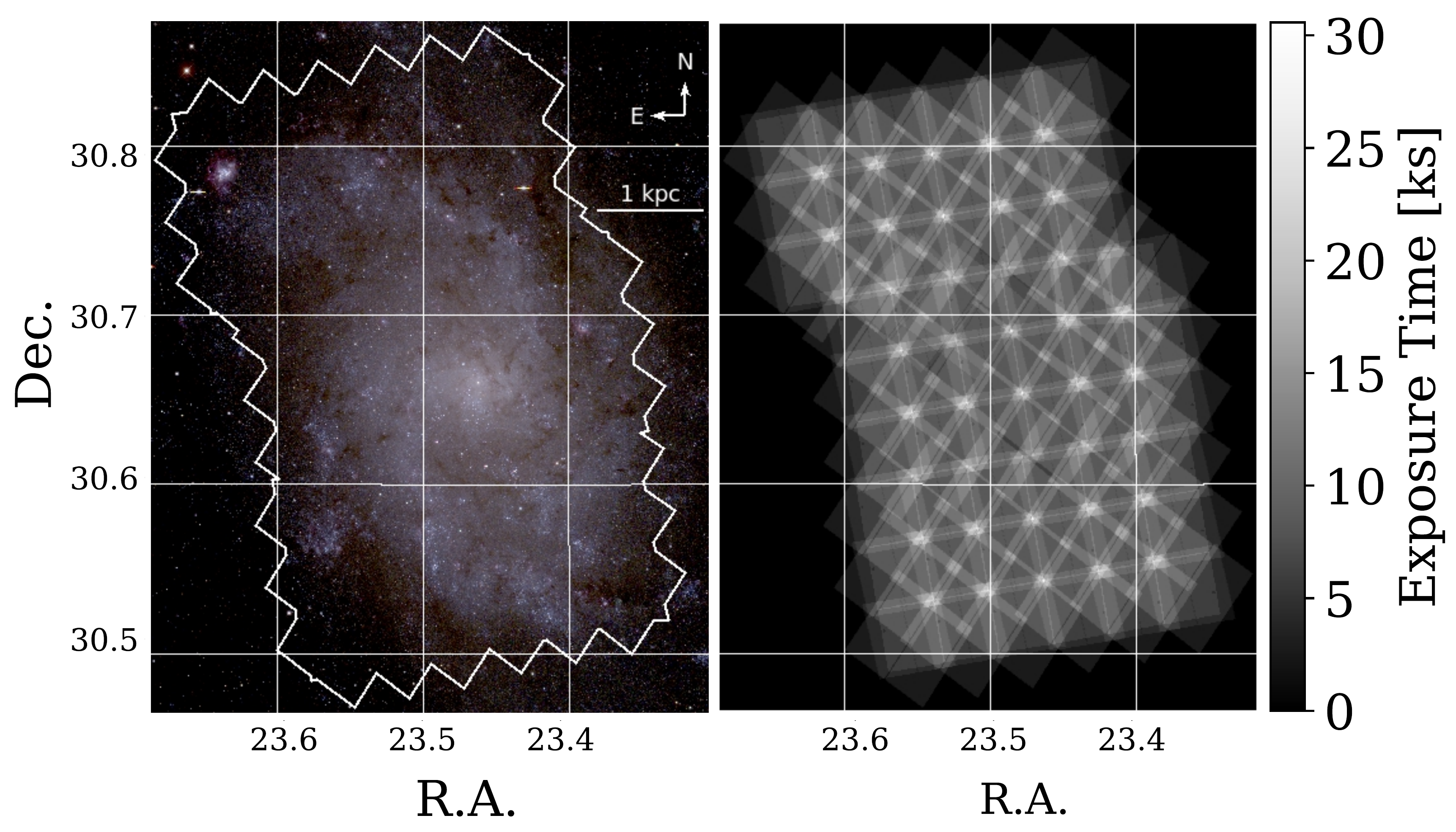}
\caption{\textbf{left:} The PHATTER survey footprint overlaid on a color optical image of M33 \citep[generated with R, V, and B band images from the Local Group Galaxy Survey;][]{Massey2006}. The outline was generated with the ACS exposure map from the PHATTER survey and includes all regions for which star formation histories were calculated for this paper. The full survey outline was divided into 2005 individual regions that measure 24$^{\prime \prime}$ on a side, using a grid of constant R.A. and Dec. \textbf{right:} Image showing the ACS exposure map from the PHATTER survey. The exposure map matches the area of the PHATTER survey outline shown in the left panel of this figure. The exposure map is color coded by total exposure time, measured in ks.}
\label{fig:phatter_outline}
\end{figure*}

We derive the spatially-resolved recent star formation history of M33 using optical photometry from the PHATTER survey\footnote{\dataset[PHATTER Data, DOI: 10.17909/t9-ksyp-na40]{https://doi.org/10.17909/t9-ksyp-na40}}. The optical-only PHATTER survey covers a projected area of $\sim300$ arcmin$^{2}$, or $\sim$38 kpc$^{2}$ de-projected, in the center of the star forming disk of M33, as shown in Figure \ref{fig:phatter_outline}. While the six-band PHATTER photometry catalog contains over 22 million individual stars and includes six filter photometry from the near-infrared to near-ultraviolet, as discussed in detail in \citet{Williams2021} the data we use in our analysis comes from the ACS (optical; F475W and F814W) data alone. This catalog from the optical-only imaging data contains $\sim$25 million stars total with $\sim$15 million of these having reliable measurements (i.e., stars that meet our stringent ``gst'' quality cuts which are described in Section \ref{sec:photometry}) in both the F475W and F814W bands. The use of the optical-only catalog was critical for creating the artificial star tests (ASTs) in a reasonable time and creating a catalog of homogeneous exposure times across the survey footprint. We compute the SFH for all of the area within the PHATTER survey outline indicated in Figure \ref{fig:phatter_outline}.

\subsection{Photometry and Catalogs}\label{sec:photometry}
We use optical photometry in the F475W and F814W bands for our measurements. We use the ``gst'' catalogs, which contain ``good stars'' that pass cuts related to their signal to noise, crowding, and sharpness as described in \citet{Williams2021}. Specifically, all stars in the ``gst'' catalog have S/N$>$4, sharpness$^{2}<$0.2, and crowding$<$2.25 in both bands. These cuts are designed to leave only the stars with the highest quality photometry for our measurements. For our analysis, we divided the ``gst'' photometric catalog into 24$^{\prime \prime}$ square regions (roughly 100 pc on a side), along lines of constant R.A. and Dec., producing 2005 regions across the survey area. The mean number of ``gst'' stars in our analysis regions is $\sim$4000, and the mean number of ``st'' stars (stars with S/N$\geq$4) is $\sim$13,000.

\subsection{Artificial Star Tests}\label{asts}
It can be challenging to accurately measure the magnitudes of stars in crowded regions. Faint stars can blend together, biasing the measured flux from other stars within the same region, such that magnitudes of stars in regions of high stellar density will have more biased, noisier measured flux. We quantify the degree of bias, noise, and completeness in our photometric measurements using ASTs.

We repeatedly insert artificial stars into an image and then run the image through our standard photometric pipeline, as described in \citet{Williams2021}. The measured difference between the input and recovered magnitude of these stars quantifies the effect of crowding in that region. The same test can be used to set the completeness. We calculate the magnitude at which 50\% of the artificial stars are recovered and use this as our limiting magnitude when we measure the SFH in each region. Because these effects are sensitive to both stellar flux and stellar density, we want to repeat this process for stars across the CMD for a range of local stellar densities.

Rather than fully populating each pixel with the required number of artificial stars, we take advantage of the fact that the photometric effects of crowding are primarily density-dependent. We can combine ASTs from across the galaxy, as long as the background stellar density is similar. We quantify local density using the number of stars in the ``st'' catalog (stars with S/N$\geq$4) per square arcsecond. We use the ``st'' rather than ``gst'' catalog to better account for the real stellar density in the region, not just the density of stars that meet our stringent ``gst'' criteria. We grouped artificial star tests by stellar density into 16 stellar density bins to increase the number of tests that can be used for each of the SFH regions. For each of the $\sim$2000 regions in our SFH analysis, we used a set of ASTs that were generated for regions of similar stellar density. The number of ASTs and the 50\% completeness limits for each stellar density bin in Table \ref{completeness_limits}. The 50\% completeness limits correspond to an old main sequence turn off (MSTO) age of about log(age/yr)=9.5 at the faint end and about log(age/yr)=9.2 at the bright end. This suggests that our recent ($\lesssim$630 Myr) SFH maps should be complete, even in the most crowded regions.

\begin{deluxetable*}{ccccc}
\tablecaption{50\% Completeness Limits \label{completeness_limits}}
\tablehead{
\colhead{Stellar Density} &
\colhead{m$_{F475W}$} &  
\colhead{m$_{F814W}$} &
\colhead{Number of ASTs} &
\colhead{Mean Dist. from Center}\\
\colhead{[stars arcsec$^{-2}$]} &
\colhead{[mag]} &  
\colhead{[mag]} &
\colhead{used} &
\colhead{of M33 [kpc]}
}
\startdata
0$-$10 & 27.6 & 26.7 & 53031 & 2.6\\
10$-$12 & 27.6 & 26.7 & 39977 & 2.7\\
12$-$14 & 27.6 & 26.7 & 57711 & 2.6\\
14$-$16 & 27.5 & 26.6 & 59811 & 2.4\\
16$-$18 & 27.5 & 26.6 & 88746 & 2.3\\
18$-$20 & 27.5 & 26.6 & 102572 & 2.1\\
20$-$22 & 27.5 & 26.5 & 163183 & 2.0\\
22$-$24 & 27.4 & 26.4 & 194591 & 1.8\\
24$-$26 & 27.4 & 26.4 & 169226 & 1.7\\
26$-$28 & 27.3 & 26.3 & 142455 & 1.6\\
28$-$30 & 27.2 & 26.2 & 152861 & 1.3\\
30$-$32 & 27.2 & 26.1 & 145009 & 1.1\\
32$-$34 & 27.1 & 26.0 & 108324 & 0.9\\
34$-$38 & 26.9 & 25.6 & 98184 & 0.6\\
38$-$42 & 26.5 & 25.2 & 54982 & 0.3\\
42$-$45 & 26.3 & 24.9 & 6096 & 0.2 \\
\enddata
\tablecomments{Table listing 50\% completeness limits used in our SFH calculations, grouped by stellar density. Column 1 indicates the range of stellar densities for which the given 50\% completeness limits apply. Columns 2 and 3 list the 50\% completeness limits in F475W and F814W, respectively. In column 4 we include the number of ASTs used for each stellar density bin. In column 5 we list the mean distance from the center of M33 for analysis regions of the given density bin. We adopt the center of M33 from the FUV D25 ellipse from \citet{GildePaz2007}.}
\end{deluxetable*}

\section{Derivation of the SFHs}\label{sec:derivation_sfhs}
We use the CMD-fitting code \verb!MATCH! to derive the SFHs \citep{Dolphin2002}. \verb!MATCH! finds the combination of single-burst stellar populations that best reproduces the observed CMD. \verb!MATCH! incorporates photometric errors and completeness that are measured in the ASTs (see Section \ref{asts}) and fits for dust extinction. The best fit SFH is then determined using a Poisson maximum likelihood technique. For more details, see \citet{Dolphin2002} and \citet{Lewis}.

Because our focus is on the recent SFH only, we exclude regions of the CMD occupied by older, redder stars. The atmospheres and evolution of main sequence stars are generally better understood than later evolutionary stages, making the SFH recovery for young ages reasonably secure. We exclude the area of the CMD dominated by these older ages, such as the vast majority of red giant branch and red clump stars, with both F475W magnitude greater than 21 and F475W-F814W color greater than 1.25. 

Our choice of exclusion region purposefully includes main sequence stars and young red and blue He burners in our fits. Red and blue He burners are useful for age dating at young ages \citep[e.g.,][]{McQuinn2012}. We chose the blue edge of our exclusion region (at F475W$-$F814W=1.25) with the goal of excluding red giant branch and red clump stars, without also excluding the most reddened young stars. If we moved our exclusion region too far towards the blue we would risk excluding the young, reddened population from our fits, which would potentially bias the results towards recovering star formation in the affected age ranges only in regions with lower-extinction values.

The average number of stars we fit on the CMD for a given analysis region is $\sim$4000, with a standard deviation of $\sim$1500 stars. The \verb!MATCH! fitting of a given region is essentially finding the intrinsic distribution of young stars and dust that reproduces the luminosity function and color distribution of the main sequence, given photometric selection, noise, and biases from the ASTs. 

In our fits we define the input the age bins, metallicity, distance, extinction, binary fraction, and choice of initial mass function (IMF). Where possible, we adopt the same model parameters that \citet{Lewis} used to fit the SFHs in M31 to make our results directly comparable. We set a fixed distance modulus of 24.67 \citep{deGrijs2014}, use a binary fraction of 0.35, and draw the mass of the secondary from a uniform distribution and Kroupa IMF \citep{Kroupa2001}. We use the same set of age bins as \citet{Lewis}, which span from 6.6 to 10.15 log(years) with a spacing of 0.1 dex, with the last bin spanning from log(years)=9.9$-$10.15. The youngest age of the isochrones used in this work is 4 Myr. We calculate the stellar mass formed in the youngest time bin (between 6.6 and 6.7 log(years) ) for each spatial region, and then derive the SFR that extends to the present day by dividing that stellar mass by the total time between the upper bin edge (log(years)=6.7) and the present day (0 Myr). The fine time resolution at young ages, within the last 25 Myr or so, results in high uncertainties in our SFR measurements, due to the small number of stars of these ages on the CMDs and theoretical uncertainties. Users of the SFH measurements presented in this paper may combine time bins in cases where they would prefer lower resolution with smaller errors.

We adopt metallicity constraints for the stellar model set used. Although main sequence evolutionary models are better constrained than other phases, there are differences between model sets, mostly at high mass. We perform our calculations using both the Padova \citep{Marigo2008} isochrones with updated AGB tracks \citep{Girardi2010}. The Padova models have a [M/H] range of [-2.3,0.1] with a resolution of 0.1 dex. We set limits on the allowed [M/H] values for the oldest and youngest time bins, allowing ranges of [-2.3,-0.9] and [-0.4,0.1], and specify that the mean metallicity may not decrease with time. The MIST models have an allowed metallicity, [M/H], range of [-2.0,0.5] and we use set the allowed [M/H] values for the oldest and youngest time bins, allowing ranges of [-2.0,-0.9] and [-0.4,0.5], respectively.

To generate our SFHs, we must generate model CMDs with stars of various ages and metallicities distributed over a two dimensional parameter space of age and metallicity. While age-metallicity degeneracy is not as strong for main sequence stars, the metallicity will strongly effect the color of the main sequence \citep[e.g.,][]{Gallart2005}. M33 has a well-established stellar and gas-phase metallicity gradient \citep{Cioni2009,Magrini2009,Magrini2010,ToribioSanCipriano2016,Lin2017} and we explore the effects of metallicity on our results in Section \ref{sec:results_metallicity}. 

\subsection{Choice of Region Size}
To generate the spatially resolved recent star formation history of M33, we use a region size of $\sim24^{\prime \prime}$ that is approximately 100 pc on a side. This region size was chosen to match the size of regions that were used to measure the spatially resolved recent star formation history of M31 to make our results directly comparable. As described by \citet{Lewis}, the 100 pc region size was chosen because it bridges the gap between previous work on Galactic pc scale star formation and star formation in more distant galaxies on kpc scales. In addition to scientific interest, the region size was chosen to provide a similar number of stars for the CMD for each individual region and to fit within the computational constraints of our analysis. Given the comparable distance of M33 to M31 and the analogous survey design, we chose to use the same region size for our analysis, which also eases comparison between the M33 and M31 results.

We generate our grid of analysis regions along lines of constant R.A. and Dec. Because of this, along the edges of the PHATTER survey footprint, some of our regions have partial coverage in the optical-only catalogs. We calculate the area of each SFH region covered by the optical-only PHATTER survey and include that value in units of [arcsec$^{2}$] in Tables \ref{table:col_info} and \ref{table:big_preview}. We use the de-projected area of each SFH region covered by the optical-only PHATTER catalog to generate figures where the SFR is shown in units of M$_{\odot}$ yr$^{-1}$ kpc$^{-2}$.

\subsection{Extinction}\label{sec:extinction}
M33 is lower metallicity and more face on (lower inclination) than M31 and thus has only moderate dust extinction \citep{Corbelli2003}. It does, however, have dust that needs to be accounted for when modeling the CMD. We include dust in our CMD fitting by including two parameters that describe the dust in each SFH region: foreground extinction ($A_{V}$), which reddens all the stars uniformly, and differential extinction in star forming regions ($dA_{V}$). We employ the same method used by \citet{Lewis} to recover extinction in the spatially resolved recent SFH measurement using the PHAT data in M31 in an effort to make our measurements directly comparable. This extinction model is adequate for young stars (age$<$1 Gyr), which are confined to the plane and tend to experience a roughly tophat differential reddening distribution \citep[e.g.,][]{Dolphin2003,Weisz2014}. Older stellar populations require more complicated fitting for extinction which follows a log-normal distribution \citep{Dalcanton2015,Williams2017}. In \verb!MATCH!, extinction is applied to all stars in a uniform distribution between $A_{V}$ and $A_{V} + dA_{V}$. Thus, $A_{V}$ shifts the entire CMD fainter and red-ward, whereas $dA_{V}$ broadens and dims the main sequence. We briefly describe the method we use for fitting $A_{V}$ and $dA_{V}$ below.

We determine the best values for $A_{V}$ and $dA_{V}$ by running our SFH calculations multiple times over a grid of $A_{V}$ and $dA_{V}$. We limit our $A_{V}$ values to a range of [0.0,1.5] and our $dA_{V}$ values to a range of [0.0,2.5]. First, we run the SFH calculations over a grid with an initial spacing of 0.2 in both values. We calculate the likelihood of each ($A_{V}$,$dA_{V}$) pair and then choose the pair with the highest likelihood value. Then we create a finer grid with 0.05 spacing in both values within 2$\sigma$ of the best ($A_{V}$,$dA_{V}$) pair and repeat our maximum likelihood technique to identify the best fit ($A_{V}$,$dA_{V}$) for the region. This method is described in more detail in Appendix B of \citet{Lewis}.

\subsection{Uncertainties}\label{sec:uncertainties}
Uncertainties in our measurement of the SFH of M33 come from a combination of random and systematic uncertainties, and dust. 

The random uncertainties were calculated using a hybrid Monte Carlo (MC) process \citep{Duane1987} implemented as part of the \verb!MATCH! software package \citep{Dolphin2013}. These random uncertainties scale with the number of stars on the CMD for a given region, with more sparsely populated CMDs resulting in higher random uncertainties. Rather than generating SFHs on a grid like we do for the $A_{V}$, $dA_{V}$ calculations, the hybridMC routine uses a Monte Carlo routine that explores high dimensional parameter spaces more efficiently than a traditional Metropolis-Hastings Monte Carlo routine. The hybrid MC routine generates 10,000 possible SFHs for the CMD from a given analysis region and the density of the generated SFHs in parameter-space is proportional to the probability density. The errors on the SFHs can then be measured by identifying the boundaries of the region containing 68\% of the samples. This method accounts for cases where the maximum likelihood SFH has zero star formation in individual time bins, allowing the uncertainties of the SFR on those time bins to be estimated.

To assess systematic uncertainties due to our choice of stellar models, we ran our full set of SFH calculations with two sets of stellar models: the Padova models with complete AGB tracks \citep{Marigo2008,Girardi2010} and the MIST stellar models \citep{Dotter2016}. Differences between stellar models mainly stem from how they model different aspects of stellar evolution including mass loss, convection, and rotation \citep{Conroy2013}. These differences tend to be greater in post-main sequence phases of stellar evolution and for the most massive stars. 

We consider uncertainties due to dust content in each individual SFH region. We expect to underestimate the star formation rate at young ages in some spatial regions due to embedded star formation. We do not expect this underestimation to affect all spatial regions equally, due to the uneven distribution of dust across the PHATTER survey footprint. We address this uncertainty in more detail and calculate the fraction of star formation we expect to have missed due to embedded star formation in Section \ref{sec:results_extinction}.

Lastly, we acknowledge that our choice of binary fraction could be an additional source of systematic uncertainty in our SFH calculations. For consistency, we adopt the same binary fraction (0.35) used by \citet{Lewis} in deriving the recent SFH for M31 using the PHAT data. As shown in \citet{Lewis}, uncertainties introduced by choice in binary model are small compared to dust variation uncertainties.

\subsection{Reliable Age Range of SFH Measurements}\label{sec:reliability_lookback}
Our SFH measurements are not sensitive to the older stellar populations that were excluded from the red side of the CMD in our fits. Our SFHs are therefore at best only reliable back to the ages of the oldest main sequence stars analyzed in each region. However, the true age back to which our fits are reliable can vary by region due to the effects of crowding and dust extinction.

\begin{figure*}
\centering
\includegraphics[width=\textwidth]{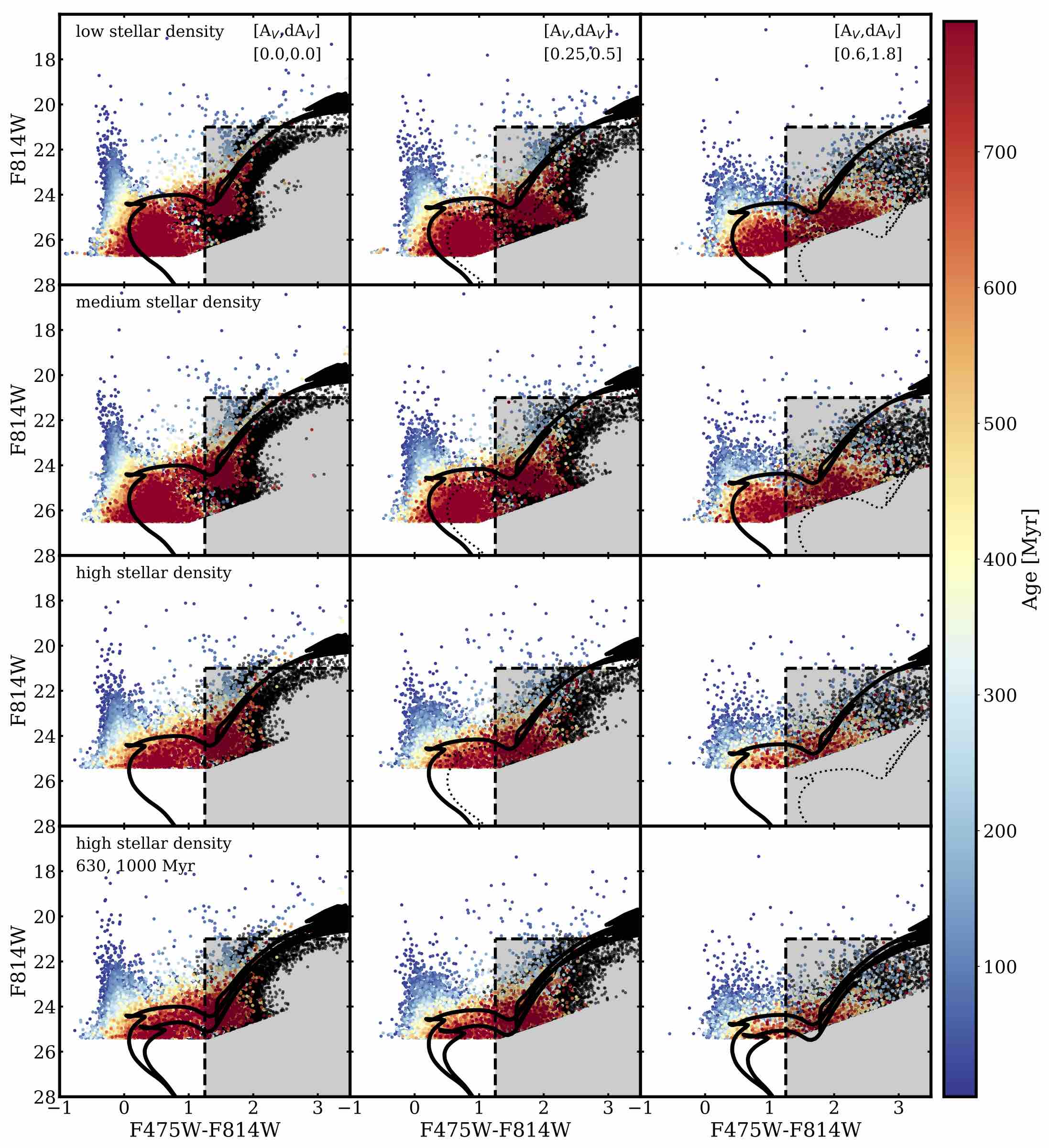}
\caption{We use artificial CMDs generated with MATCH to explore the reliability of our SFH measurements with respect to lookback time. The CMDs were all  generated for solar metallicity, with a constant SFR. Three extinction values are applied: one low ([A$_{V}$,dA$_{V}$]= [0.0,0.0]), one which is typical of our SFHs ([A$_{V}$,dA$_{V}$]= [0.25,0.5]), and one on the upper limit of our measured extinction ([A$_{V}$,dA$_{V}$]= [0.6,1.8]), shown in the left, middle, and right column, respectively. Each row shows CMDs with different stellar densities, specified in the upper left corner of the plots in the first column. For more details in the exact densities used, see Section \ref{sec:reliability_lookback}. The points on each CMD are color-coded by stellar age in Myr. All stars older than 800 Myr are plotted in black. The gray box outlines the range of colors and magnitudes excluded in our fits. In the first three rows we plot two 630 Myr isochrones reddened with the given column's A$_{V}$ (thick solid line) and $A_{V}+dA_{V}$ (thin dotted line). In the fourth row, we plot artificial CMDs for high stellar density and 630 Myr and 1 Gyr isochrones, both reddened with $A_{V}$. For further discussion, see Section \ref{sec:reliability_lookback}.}
\label{fig:fake_cmds}
\end{figure*}

To determine the sensitivity of our SFH fits as a function of lookback time, we created artificial CMDs with \verb!MATCH!, which we present in Figure \ref{fig:fake_cmds}. We created artificial CMDs with constant SFR across age bins and solar metallicity and varied the stellar density and dust extinction, which are the two main factors that would affect the reliability of our SFHs with respect to lookback time. We created artificial CMDs across a grid of stellar density and dust extinction. In the first three rows of Figure \ref{fig:fake_cmds} we show artificial CMDs for low, medium, and high stellar density regions. We repeat the artificial CMDs for the high stellar density region in the fourth row. The points on the artificial CMDs are color-coded by stellar age. All stars older than 800 Myr are plotted in black and stars younger than 800 Myr are color coded by age according to the colorbar on the right-hand side of the figure.

The low, medium, and high stellar density regions were created with stellar densities of 0$-$10, 22$-$24, and 34$-$38 stars per square arcsecond, respectively. To generate artificial CMDs at each stellar density, we used the completeness limits measured for each stellar density bin and used ASTs for the given stellar density. Our artificial CMDs were created with three levels of dust extinction: no dust extinction ([A$_{V}$,dA$_{V}$]=[0.0,0.0]), median dust extinction ([A$_{V}$,dA$_{V}$]=[0.25,0.5]), and high dust extinction ([A$_{V}$,dA$_{V}$]=[0.6,1.8]). Median and high dust extinction are defined from the range of A$_{V}$ and dA$_{V}$ values we measured in our SFHs.

In the first three rows of Figure \ref{fig:fake_cmds}, we plot 630 Myr isochrones reddened with the $A_{V}$ specified for each column with a thick black line. We also include a thinner, dotted line showing the same age isochrone reddened with $A_{V} + dA_{V}$, to show the most extreme reddening applied to stars in our fits. In the fourth row, we plot a 630 Myr isochrone reddened with the $A_{V}$ specified for each column and a 1 Gyr isochrone with the same reddening applied. 

We find that at low stellar density and extinction our photometry depth allows us to observe main sequence stars back to about 1 Gyr. However, when the depth is limited by crowding in the high stellar density regions, the main sequence turn-off for stars older than $\sim$800 Myr falls at or near the magnitude limit, meaning that our SFH measurements are not reliable at these ages. We thus adopt a conservative age limit of 630 Myr for our scientific analysis in this paper, although lower density and less-extincted regions may have reliability back to older ages. This reliability limit is older than the one adopted by \citet{Lewis} in the spatially resolved recent SFH of M31 (400 Myr). This difference is caused by the higher crowding of M31, due to its more edge-on inclination. Higher crowding results in brighter magnitude limits, resulting in SFH measurements that are reliable back to younger ages.

We include figures showing CMDs with our data for three regions (low, medium, and high stellar density) and compare these CMDs with model CMDs in Appendix \ref{sec:cmd_residuals}.

We ran tests where known SFH and extinction were used to create synthetic CMDs, from which the SFH and extinction were fit. We ran these tests with realistic SFHs and with single bursts of star formation. We discuss these tests and their implications in detail in Appendix \ref{sec:sfh_recovery_tests}.

\section{Results}\label{sec:results}
In this section we present the integrated and spatially resolved SFHs for M33. We present the full SFH results for each spatial region in Tables \ref{table:col_info} and \ref{table:big_preview}, where we present column descriptions and the full SFR measurements and the coordinates of the axes of each individual region. We include the SFH for the first six spatial regions in Table \ref{table:big_preview} and include the full table for all spatial regions in the digital version of this paper. Unless otherwise specified, figures and tables showing SFH results were created with the Padova model measurements.

\begin{figure*}
\centering
\includegraphics[width=0.95\textwidth]{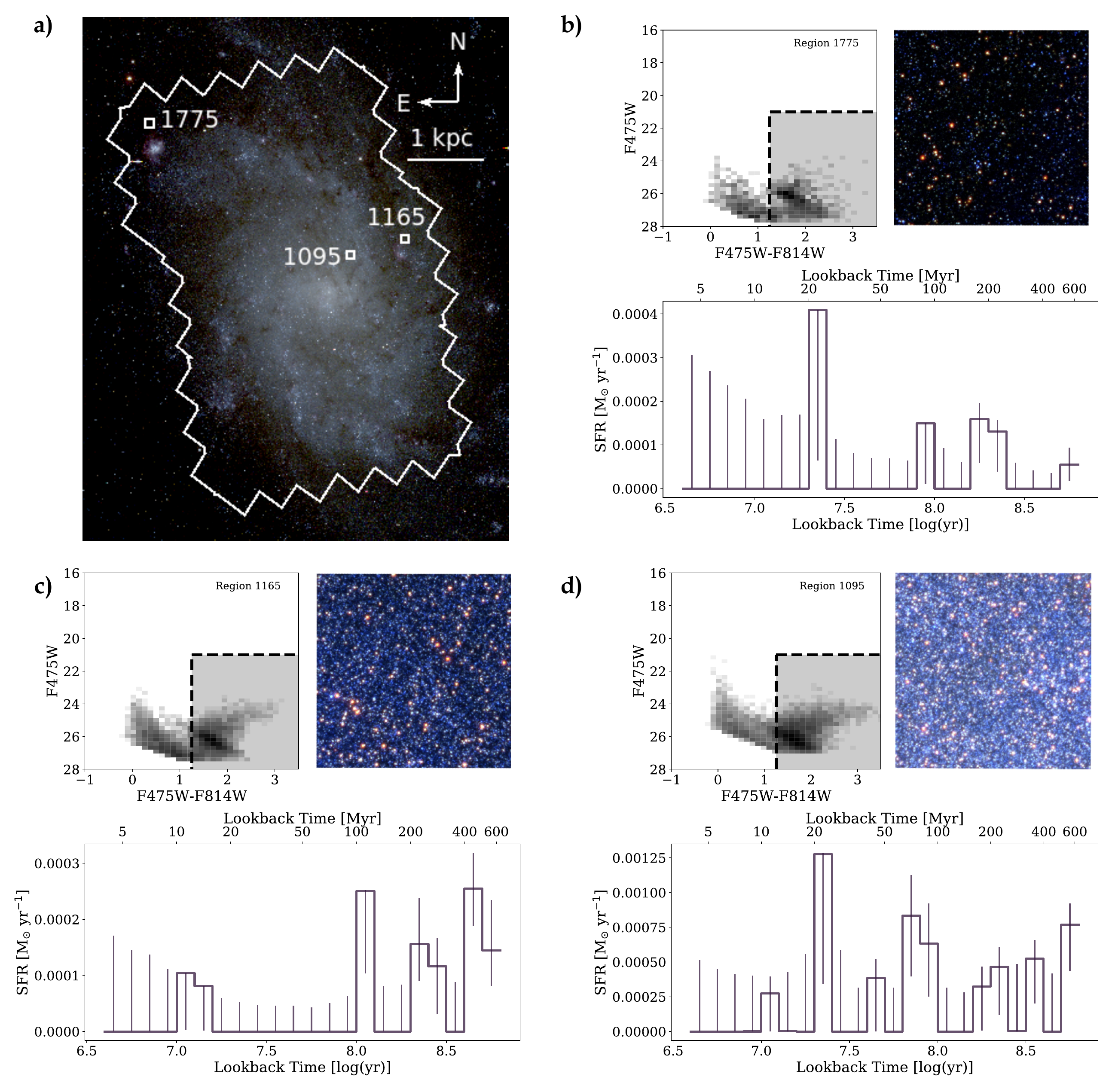}
\caption{\textbf{(a)} We provide a color image of M33 with the PHATTER survey outline and the location of three regions marked for context. Region 1775 is a low stellar density region, 1165 is a medium stellar density region, and 1095 is a high stellar density region. The image is composed of B, V, and R filter images \citep{Massey2006}. \textbf{(b)} Panel b includes three sub-figures for region 1775, a low stellar density region. The upper left plot shows a CMD for the stars in that region. The gray area outlined with a dashed black line indicates the excluded region that was not included in our SFH measurements. The upper right plot is a color image of region 1775, which measures 24$^{\prime \prime}$ on a side, roughly 100 pc at the distance of M33. This image was created using the PHATTER survey mosaic images in the following bands: F160W (red), F814W (green) and F475W (blue). The star formation history over the last 630 Myr for this region is shown in the bottom plot, with random errors plotted. \textbf{(c)} Panel c shows the same set of three figures for region 1165, a medium stellar density region. Note how the stellar density in the finder image is visually higher, and the CMD is more filled out. The SFR measurements are also higher, indicating more stellar mass being formed in this region over the last 630 Myr. \textbf{(d)} Panel d shows the same three figures for region 1095, a high stellar density region near the center of M33. This region has the most stars in its CMD of the three regions illustrated in this figure and also shows the highest SFR measurements over the past 630 Myr.}
\label{fig:density_sfh_cmd}
\end{figure*}

In Figure \ref{fig:density_sfh_cmd}, we explore the effect of stellar density on our SFH measurements for individual regions. In panel (a) we include a color image of M33 with the PHATTER survey footprint outlined and the location of three of our analysis regions outlined and labeled with their region id numbers. These regions were chosen for their varying stellar densities. Region 1775 is a lower stellar density region ($\sim$7 stars arcsec$^{-2}$), region 1165 is a medium stellar density region ($\sim$24 stars arcsec$^{-2}$), and region 1095 is a high stellar density region ($\sim$34 stars arcsec$^{-2}$). For each region, we include a CMD of the ``gst'' stars in that region, with the region of the CMD excluded from our fits plotted as a gray box. Stars that lie outside of the excluded region were used to measure the SFH in that region. In panels (b, c, d) we include a color composite image created with the PHATTER survey mosaics (red=F160W, green=F814W, blue=F475W). The SFH plotted for each region includes a histogram showing the best-fit SFR for each time bin, with random uncertainties. The uncertainties presented for each region are typical of those for regions of similar stellar density. While the increased stellar density can affect our completeness limits, the increased number of stars being fit on the CMD results in smaller error bars on our SFR measurements.

\begin{figure*}
\centering
\includegraphics[width=0.95\textwidth]{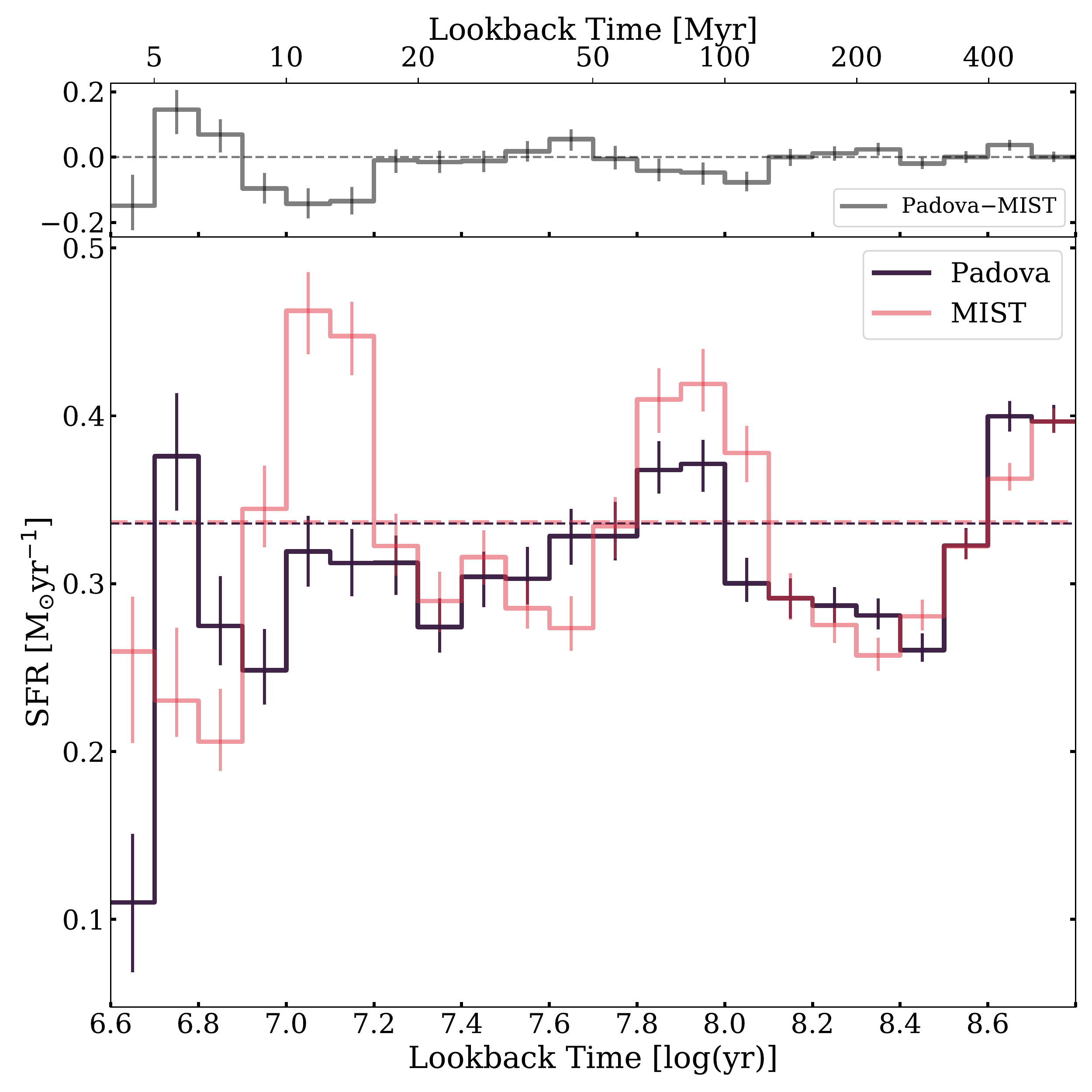}
\caption{The total SFH for the PHATTER survey area, which covers $\sim$38 square kpc, de-projected. This figure combines spatial bins in each 0.1 log(year) time bin, using the full time resolution of our fits. \textbf{top:} The difference between the best-fit SFH measurements using the Padova and MIST stellar models is plotted as a function of time. The dashed horizontal line marks a difference of 0. \textbf{bottom:} The black histogram represents the SFH when we use the Padova stellar models in our CMD fitting and the pink histogram shows the SFH when we use the MIST stellar models. The error bars represent the random uncertainties. The horizontal dashed lines mark the mean SFR value for each set of stellar models, the mean SFR values agree so closely, they are difficult to distinguish on the plot. The two SFH fits illustrate systematic uncertainties in our measurements. The models agree very well up to about 20 Myr ago. Differences at younger ages reflect differences between how the two model sets treat the most massive, young stars. For a full discussion of the differences between these two SFH measurements and systematic uncertainties in our SFH measurements, see Section \ref{sec:uncertainties}. The SFR values and error bars plotted for the Padova stellar models are listed in Table \ref{table:total_sfh}. Due to the extent of our survey area, we expect to have measured roughly 45\% of the SF in M33, which we determined by comparing to the fraction of total FUV flux within the PHATTER survey footprint.}
\label{fig:total_sfh}
\end{figure*}

The co-added SFH measurements for all sub-regions reveal the integrated recent SFH of M33 within the PHATTER footprint. We add the stellar mass formed within all spatial bins for the same time bin. The integrated SFH within the PHATTER footprint obtained with the Padova models is listed in Table \ref{table:total_sfh}. We list the start and end of each time bin in both log(years) and Myr and include uncertainties, which include both random and dust uncertainties. We plot the spatially integrated SFH in Figure \ref{fig:total_sfh}, where we present the best-fit SFH obtained with both the Padova and MIST models as a function of time. The dark histogram represents the Padova models and the lighter pink histogram represents the MIST models. 

The horizontal dashed lines in the lower panel of Figure \ref{fig:total_sfh} (which overlap) indicate the mean SFR within the last 630 Myr ($\sim0.34$ M$_{\odot}$ yr$^{-1}$) derived for the PHATTER survey area for the MIST and Padova models. While their SFR measurements within the youngest time bins differ, the mean SFR within the last 630 Myr measured with the two models differs by $<$1\%.

The MIST and Padova models agree within their random uncertainties for most time bins older than about 20 Myr, indicating that random uncertainties dominate the errors on the SFR at these ages. The largest systematic differences in our SFR measurements are in the youngest time bins, where models for the most massive stars generate most of the uncertainty. At the youngest ages ($<$15 Myr), the model sets differ by up to a factor of two in the measured SFR.   

\begin{figure*}
\centering
\includegraphics[width=0.95\textwidth]{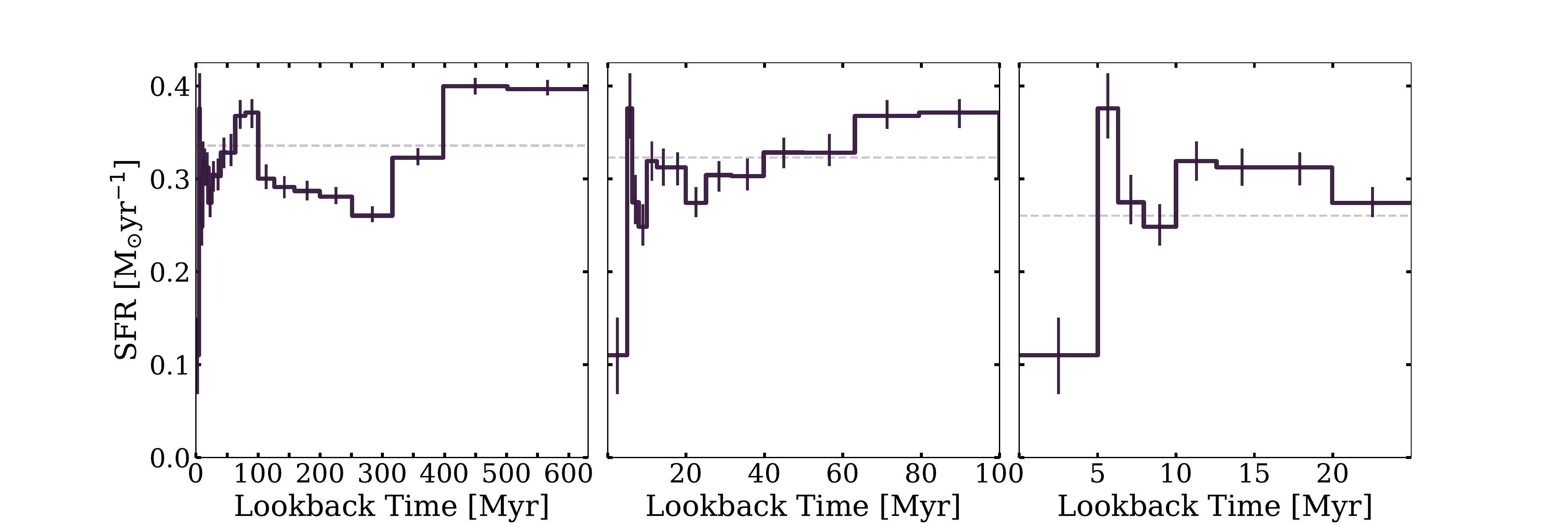}
\caption{The total SFH for the PHATTER survey area from the Padova stellar models. These figures use a linear time scale on the x-axis to illustrate the resolution of our fits at different time scales. The left panel shows the SFH back to 630 Myr ago, the middle panel shows the SFH back to 100 Myr ago, and the right panel shows the SFH back to 25 Myr ago. The dashed horizontal line in each panel represents the mean SFR over the plotted timescale. The spatially-integrated SFR values and error bars plotted are listed in Table \ref{table:total_sfh}.}
\label{fig:linear_sfh}
\end{figure*}

We also present the SFH (for the Padova models) plotted linearly in Figure \ref{fig:linear_sfh}, to better represent the resolution of the youngest time bins. We find that the SFR of M33 is fairly constant over the past $\sim$500 Myr. Both the Padova and MIST model measurements suggest some recent elevation in the SFR, but they place it at slightly different times and amplitudes. The Padova models find two peaks in the SFR within the past 100 Myr, around 5$-$10 Myr and 70$-$100 Myr ago, with an elevation of a few times $10^{-2}$ M$_{\odot}$ yr$^{-1}$ above the mean SFR over the past 630 Myr.

\subsection{SFH Maps}
In Figure \ref{fig:sfr_log_bins_map}, we show the spatially-resolved SFH, plotting maps of the SFR measured in each spatial bin within the time ranges specified in the upper left corner of each panel. At the youngest ages (upper left), we have grouped the time bins together into bins of roughly 25 Myr to create smoother maps for our analysis. Each analysis region is represented as a pixel, with the color of the pixel corresponding to the best-fit SFR during the time span specified. Regions plotted in black have a measured SFR of below $10^{-4}$ M$_{\odot}$ yr$^{-1}$, comparable to the typical uncertainty in low SFR regions. There are clear time-dependent changes in the pattern of SF, which we discuss in Section \ref{sec:discussion}.

We note that towards the older bins in Figure \ref{fig:sfr_log_bins_map}, the central region of the galaxy does not have detectable star formation, suggesting that the significant star formation in the galaxy center may have started only $\sim$100 Myr ago. Our data in this region are sensitive to these stellar ages, so this timescale for the star formation at the galaxy center may be robust. However, this region also shows significant dust content (see Figure \ref{fig:dAv_24micron}), with high values of $dA_{V}$, suggesting that we may not be seeing the full characteristics of the young populations in this region. 

We also present the integrated SFH over the last $\sim$630 Myr in Figure \ref{fig:total_sfr_map}. To generate this figure, we summed the total stellar mass formed in each sub-region over all time bins back to 631 Myr. We find the integrated SFH shows clear spiral structure with a prominent two arm spiral that looks quite different from the more flocculent structure visible in an optical image. In addition there is a noticeable bar structure. We return to these points in detail in Section \ref{sec:discussion}.

\subsection{SFR Over Different Timescales}\label{sfr_timescales}
It is instructive to calculate the integrated SFR over various timescales (0$-$10, 0$-$100, 0$-$500, and 0$-$630 Myr) for comparison with different extragalactic SFR indicators, which we include in Table \ref{table:sfr_timescales}. We also include the SFR over these timescales for each individual region in Table \ref{table:big_preview}.

The average SFR within the PHATTER footprint within the past 100 Myr is useful to compare with SFR measurements involving broadband FUV flux \citep[see][]{Kennicutt&Evans2012}. When restricted to the most recent 100 Myr, the average SFR within the PHATTER survey is 0.32$\pm$0.02 M$_{\odot}$ yr$^{-1}$ or 8.4$\pm$0.5$\times$10$^{-3}$ M$_{\odot}$ yr$^{-1}$ kpc$^{-2}$. We plot the spatially-resolved SFR over the past 100 Myr in Figure \ref{fig:100Myr_FUV_comparison}, which we show alongside a FUV image of M33 \citep{GildePaz2007}.

\subsection{Recovered Extinction}\label{sec:results_extinction}
\begin{figure*}
\centering
\includegraphics[width=\textwidth]{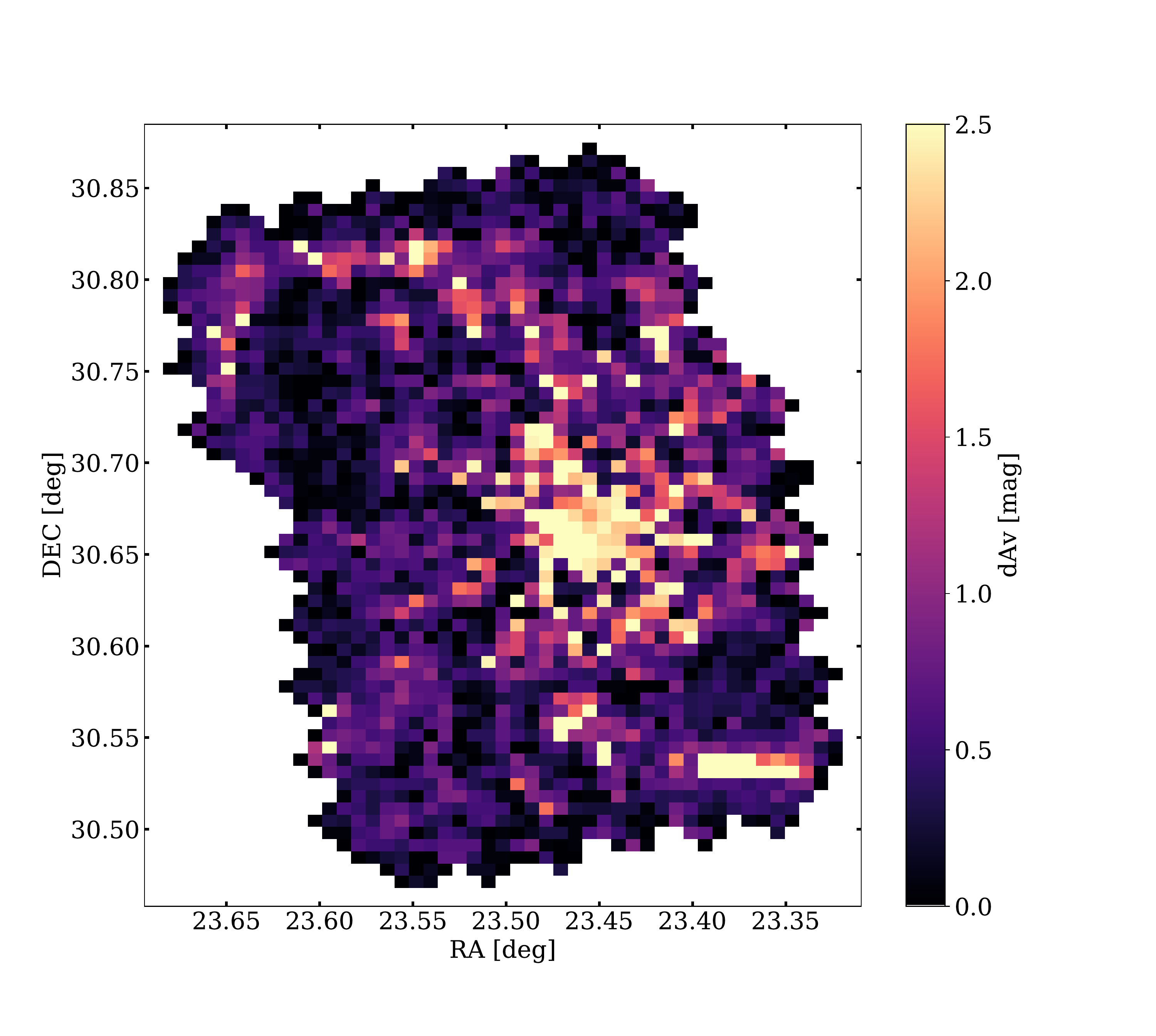}
\caption{We present a map showing out best fit differential extinction (dA$_{V}$) value for each analysis region in the PHATTER survey area. We observe that the high extinction regions follow the spiral structure of M33, suggesting we adequately recovered dust extinction in our fits. We expect that our dA$_{V}$ maps may differ slightly from maps of dust emission due to the geometry of stars and dust in individual regions (i.e., if stars are in front of or behind dust clouds from our point of view). We describe our method for measuring dA$_{V}$ in Section \ref{sec:extinction}.}
\label{fig:dAv_24micron}
\end{figure*}

We measure relatively modest foreground $A_{V}$ values across the PHATTER survey footprint with values between 0.0 and 0.5 mag. We measure a wider range in differential extinction, finding $dA_{V}$ values ranging from 0.0 to 2.5 magnitudes, with larger values of $dA_{V}$ in the center and concentrated in spiral arms. When we plot the $dA_{V}$ values for regions across the PHATTER footprint, we find that the morphology follows the galaxy's spiral structure, which suggests that we recover the intrinsic dust extinction in our fits. In Figure \ref{fig:dAv_24micron} we present a map showing the best-fit $dA_{V}$ value for each analysis region. We include the best-fit $A_{V}$ and $dA_{V}$ value for each region in Table \ref{table:big_preview}.

For all regions with high measured dust extinction ($A_{V} + dA_{V} > 2.3$), we looked at each region's CMD and measured SFH by eye to confirm that the high dust extinction wasn't a result of a failure of our fitting code. There was only one region for which the high $A_{V} + dA_{V}$ value was the result of a catastrophic failure of the code, region 1595 which contained a shredded foreground star. We physically masked the foreground star and its diffraction spikes out of the region's photometry and re-calculated its SFH. We present the corrected SFH for region 1595 in all results.

Even though our measured $A_{V}$ and $dA_{V}$ values seem to accurately recover the known foreground extinction to M33 and the expected dust morphology, we still expect to miss some star formation in M33 due to embedded star formation. The geometry of the dust and young stars -- whether the stars are in front of or behind the dust from our vantage point -- will affect our measurement of $dA_{V}$ and the star formation rate. We expect to miss some star formation in regions where the young stars are hidden behind too much dust. As shown in Figure \ref{fig:fake_cmds}, at the largest $dA_{V}$ values, some of the young stars are reddened all the way into the exclusion region of the CMD, meaning that our measurements are not sensitive to levels of dust reddening beyond $\sim A_{V}+dA_{V} \gtrsim 2.4$.

\begin{figure*}
\centering
\includegraphics[width=\textwidth]{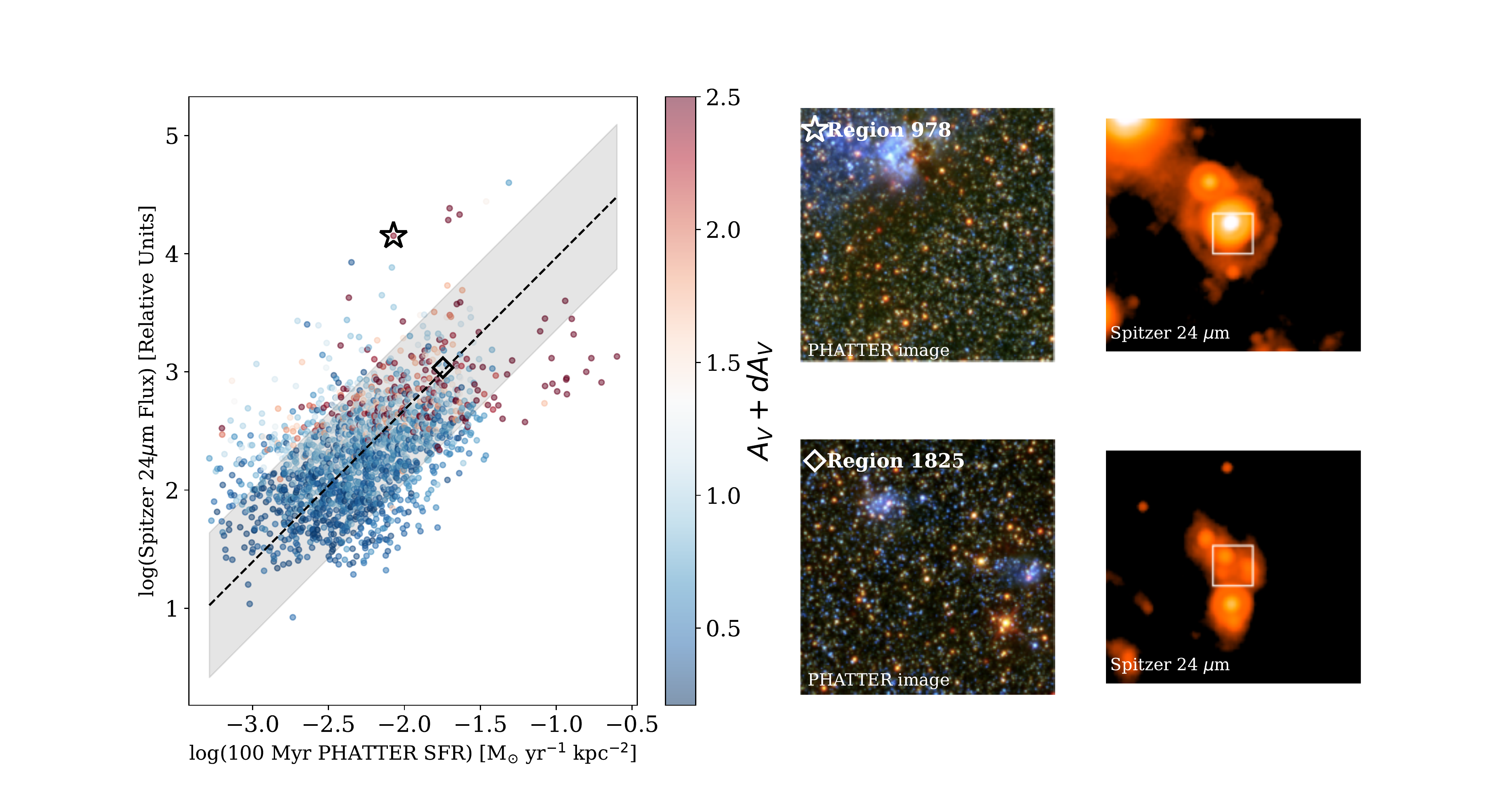}
\caption{\textbf{left:} We present a comparison between our measured 100 Myr SFR in each analysis region and its 24 $\mu$m flux measured with Spitzer, plotted in units relative to other pixels in the image \citep{Tabatabaei2007}. We plot the line of highest correlation with the black dotted line and show the 1$\sigma$ range shaded in gray. The points are color coded based on the measured $A_{V} + dA_{V}$ in that region. We expect to have under-estimated the star formation rate for regions that lie above the 1$\sigma$ region, because those regions have lower 100 Myr SFR than expected based on their 24 $\mu$m flux, suggesting that we have missed embedded star formation. \textbf{right, top row:} We present an optical color image from the PHATTER survey and a 24 $\mu$m image from Spitzer for region 978 (outlined on the left panel with a black star), where we expect to have underestimated star formation due to embedded star formation. The PHATTER image measures 24$^{\prime \prime}$ on a side and was created using the PHATTER survey mosaic images in the following bands: F160W (red), F814W (green), F475W (blue). The Spitzer 24 $\mu$m image shows a white square outlining the 24$^{\prime \prime}$ shown in the PHATTER image. For reference, the Spitzer PSF is $\sim$6$^{\prime \prime}$, or one quarter the length of the sides of the white square. Blue light from active star formation and a dust cloud are visible in the PHATTER image. In the Spitzer image, we see bright emission from the warm dust in the star forming region. Region 978's measured star formation is lower than expected given its high 24 $\mu$m flux, suggesting we may have missed star formation embedded in dust. \textbf{right, bottom row:} We present optical and 24 $\mu$m images of region 1825, which is outlined on the left panel with a black diamond. The 24 $\mu$m flux and measured 100 Myr SFR for this region agree, suggesting that we have not missed significant star formation embedded within dust.}
\label{fig:ir_sfr_comparison}
\end{figure*}

We expect that our measured 100 Myr SFR for a given region should correlate with the infrared emission from dust warmed by young stars (24 $\mu$m). To estimate the fraction of star formation that we miss due to embedded star formation, we look for regions where the 24 $\mu$m emission suggests that there is significant star formation, but our measured 100 Myr SFR is lower than expected. In Figure \ref{fig:ir_sfr_comparison} we plot our measured 100 Myr SFR for each analysis region against the 24 $\mu$m flux in that region from Spitzer \citep{Tabatabaei2007}. We re-binned the Spitzer image to match the pixel scale of our SFH maps for comparison and plot relative units of flux within the re-binned image. The point for each region is color coded by its best-fit $A_{V}+dA_{V}$. We include the line of highest correlation as a black dotted line and include the 1$\sigma$ range in gray. 

The clear correlation between our rates and the warm dust emission suggests that we do capture most of the star formation associated with dusty regions. However, we expect to have underestimated star formation in the regions that lie far above the line of highest correlation in the left panel of Figure \ref{fig:ir_sfr_comparison}, as these regions have lower 100 Myr SFR than their 24 $\mu$m flux predicts, suggesting we have missed star formation embedded in dust. 

To quantitatively look for potential embedded star formation, we visually inspected optical and 24 $\mu$m images of regions that lie above the 1$\sigma$ range in Figure \ref{fig:ir_sfr_comparison}. We generated optical finder images using the PHATTER mosaic images in the following bands: F160W (red), F814W (green), and F475W (blue). We generate IR finder images using the full-resolution Spitzer 24 $\mu$m images of M33 \citep{Tabatabaei2007} \ref{fig:ir_sfr_comparison}. In Figure \ref{fig:ir_sfr_comparison} we present optical and IR finders for two different regions: one where the 24 $\mu$m flux suggests we may be missing embedded star formation (Region 978) and one where the 24 $\mu$m flux and 100 Myr SFR agree (Region 1825). In the optical image of Region 978, we see blue light from active star formation and a dust cloud that is most visible just below the bright blue star forming region at the top of the image. In the IR image, the white square outlines the area shown in the optical image. The 24 $\mu$m flux and 100 Myr SFR for this region are plotted on the left panel, outlined in a black star. We see that the 100 Myr SFR is lower for this region than we would expect given its bright 24 $\mu$m flux, suggesting that we may be missing some star formation embedded within the dust. The 100 Myr SFR measured for region 1825, on the other hand, matches up with the region's 24 $\mu$m flux (outlined with a black diamond on the left panel), suggesting that even though warm dust is present in this region, we are not missing significant star formation.

To quantitatively estimate the fraction of star formation missed, we can look at the outliers in the left panel of Figure \ref{fig:ir_sfr_comparison}. If we assume that at least some of the outliers are indicating missed star formation (for an example of this, at more than 2 sigma above the line), we can calculate the amount the SFR in each of these outlying regions would need to be increased to move them onto the line of highest correlation. We identify 21 regions that lie above 2$\sigma$ and estimate that in these regions, we could be missing $\sim$7\% of the total star formation within the last 100 Myr due to embedded star formation. 

The results presented in this section describe the outcome of our A$_{V}$ and dA$_{V}$ fits with the Padova models. We also fit for extinction with the MIST models. When we compare the measured A$_{V}$ and dA$_{V}$, we find that the two model sets agree on recovered extinction tile to tile with a mean difference in the A$_{V}$ measurements of 0.03$\pm$0.07 mag. The mean tile to tile difference in dA$_{V}$ is 0.01$\pm$0.3 mag, suggesting that there is not an offset in the recovered dA$_{V}$ values, but that there is scatter. There is not a spatial trend across the galaxy in the difference between the extinction measurements with the Padova and MIST models.

\subsection{Recovered Metallicity}\label{sec:results_metallicity}   
Our recent star formation histories should not be strongly sensitive to metallicity because the color of the upper main sequence CMD feature that our SFHs rely on is not strongly impacted by metallicity \citep[e.g.,][]{Bressan2012}. However, there is some sensitivity from the colors of more evolved Helium burning stars, which could potentially yield a signal when combined over many measurements. We did allow some freedom in the metallicity to be fitted between -2.3$<$[M/H]$<$0.1 for the Padova models and between -2.0$<$[M/H]$<$0.5 for the MIST models. One way to determine if our SFH results have any metallicity sensitivity is to compare the slope of the best-fitting mean metallicities with radius to the slope of the known M33 gas-phase metallicity gradient, such as that measured with observations of H II regions and planetary nebulae \citep[e.g.,][]{Magrini2010}.

To make this comparison we calculated the mean metallicity recovered in our SFH fits radially across the PHATTER survey footprint. We binned the SFH regions into annuli with a width of 1 kpc from the D25 ellipse center of M33 \citep{GildePaz2007} and measured the mean metallicity in each radial bin. Due to the large uncertainties on our metallicity measurements, we find that the slope of the metallicity gradient recovered with both the Padova and MIST models is consistent with both a flat gradient and the slope of the gas-phase metallicity gradient from \citet{Magrini2010}, within 1$\sigma$ errors. Additionally, we do not see a correlation (either positive or negative) between our measured metallicity and A$_{V}$ or our measured metallicity and dA$_{V}$ when we compare tile to tile, suggesting that we are not seeing metallicity as an artifact of the metallicity-extinction degeneracy.

\section{Discussion}\label{sec:discussion}
The SFH maps reveal the time evolution of several interesting structures in the disk of M33, among these are the spiral arms and a central bar. We discuss both of these examples in detail below.

\subsection{Evolution of Spiral Structure}
\begin{figure*}
\centering
\includegraphics[width=0.95\textwidth]{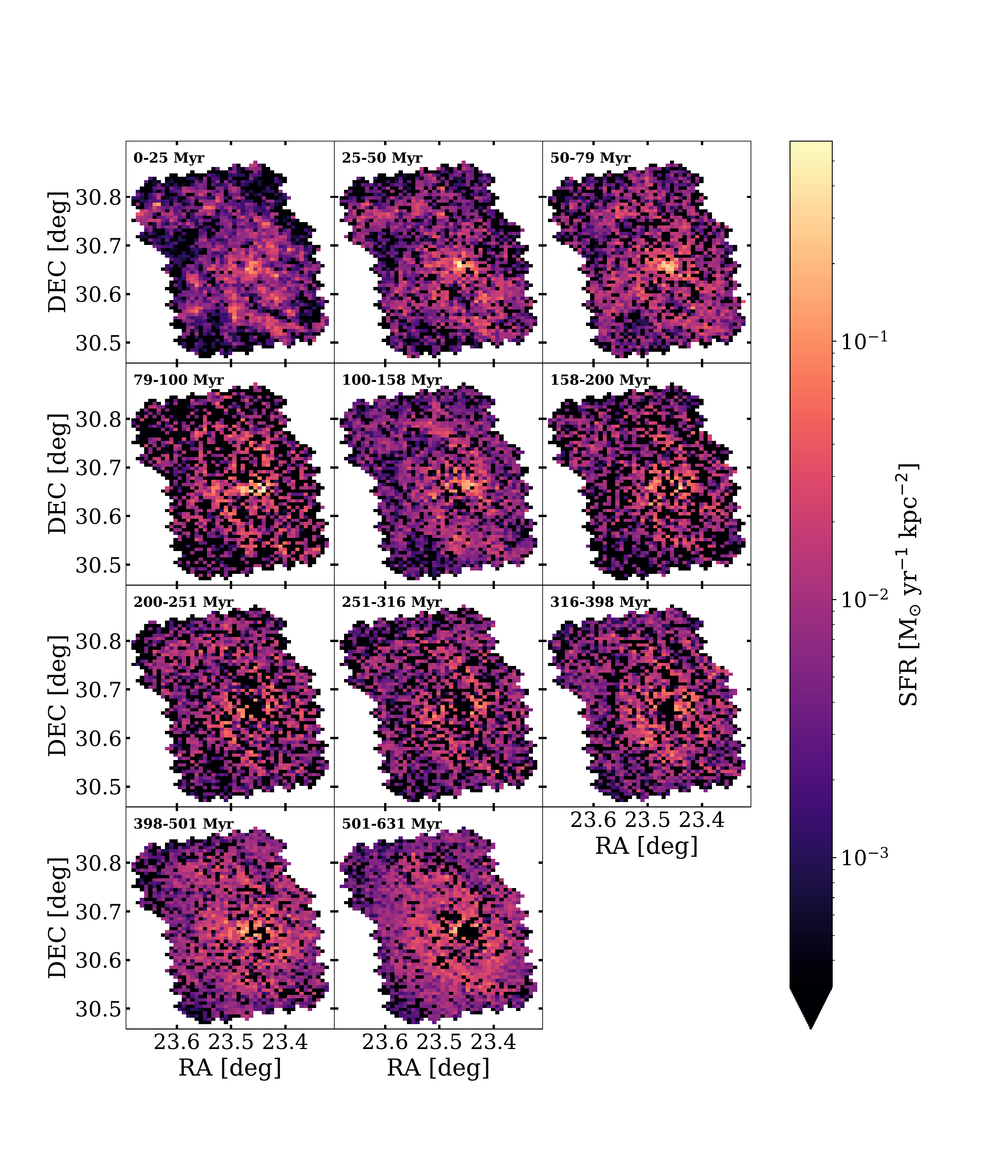}
\caption{SFH maps of M33 within the PHATTER footprint area over indicated timescales. These time bins do not reflect the full 0.1 log(year) resolution of our SFH fits, but we group the SFH up to 100 Myr ago into $\sim$25 Myr bins to more clearly show the structure of the spiral arms. The SFR values shown are the best-fit values from our fits. Each pixel represents a 100 pc by 100 pc region for which we derived the SFH using CMD-fitting. We note that the central pixels in the later time bins (older than 100 Myr) show 0 SFR, as noted in Section \ref{sec:results}. The cause for this is not yet clear, but could be due to incompleteness.}
\label{fig:sfr_log_bins_map}
\end{figure*}

\begin{figure*}
\centering
\includegraphics[width=\textwidth]{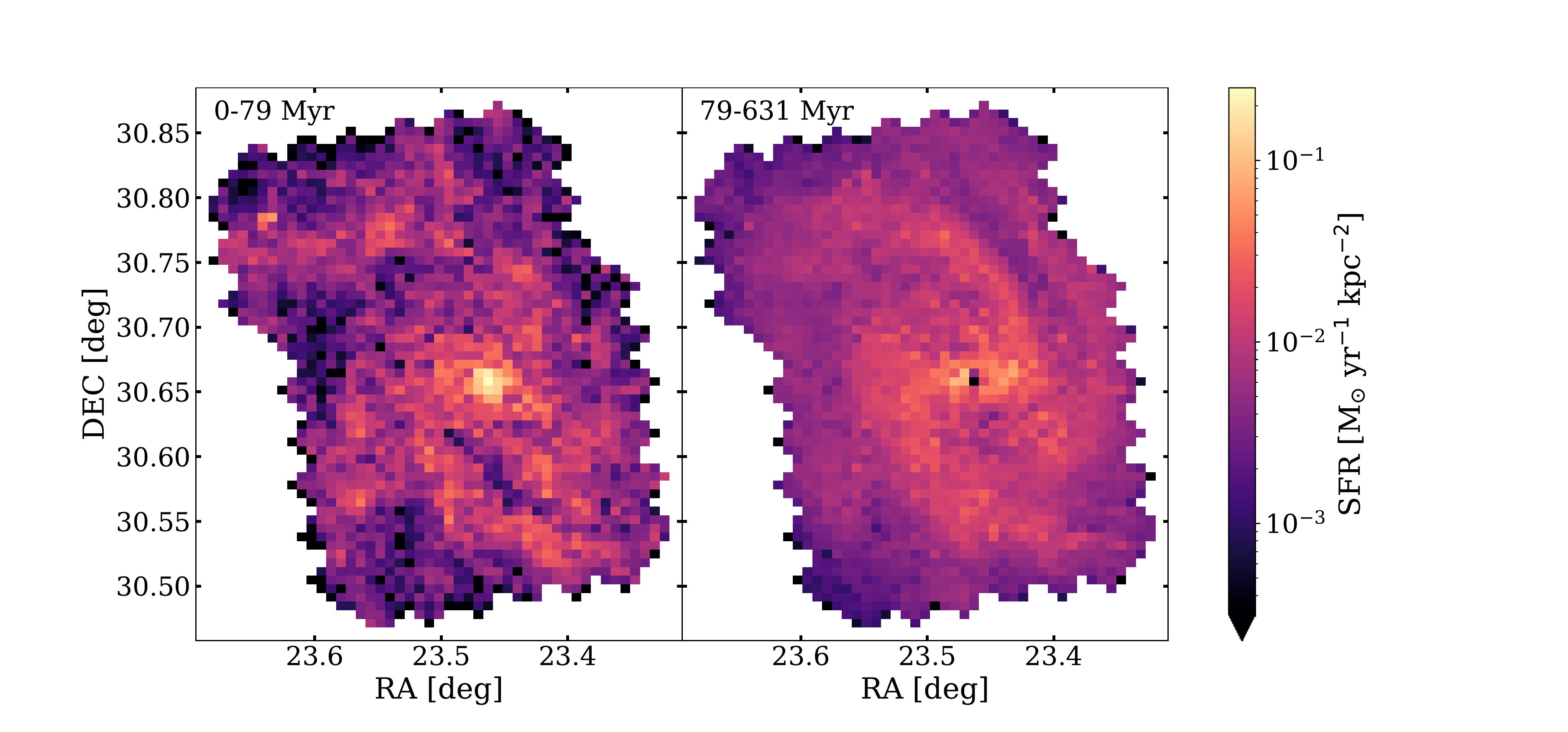}
\caption{We present the spatially resolved star formation rate in M33 before and after 79 Myr ago. The left panels combines all SFR measurements for time bins between 0 and 79 Myr ago. The right panel combines the SFR measurements for all time bins between 79 and 631 Myr ago. There is a clear difference between the spiral structure in the left and right panel, indicating a transition in the spiral structure of M33 around 79 Myr ago from a two-armed barred spiral structure (right) to the more flocculent spiral structure we observe today.}
\label{fig:stacked_sfh_map}
\end{figure*}

The time resolved structure in the SFR maps in Figure \ref{fig:sfr_log_bins_map} show clear spiral structure. However, this structure appears qualitatively different at early and late times. In the earliest time bins (top row), we observe a flocculent spiral structure between the present day and $\sim79$ Myr ago. Starting in the fourth time bin (79$-$100 Myr ago) however, we begin to see the emergence of a strong two armed spiral structure, which is evident in all of the time bins $>$100 Myr. This change in morphology can be difficult to see at the full time resolution of Figure \ref{fig:sfr_log_bins_map}. To better visualize the evolution in spiral structure and reduce the impact of random small scale fluctuations in SFR measurements, we combined the SFR measurements for all bins younger than 79 Myr (which are somewhat noisy due to the stochastic nature of SF over short time intervals) and all bins older than 79 Myr. We present these time-integrated SFH maps in Figure \ref{fig:stacked_sfh_map}, showing the pronounced difference between the structures formed by younger and older stars. 

M33 has typically been characterized as a flocculent spiral galaxy with a number of weaker spiral arms, rather than the strong arms seen in grand design spirals \citep[e.g.,][]{Humphreys&Sandage1980,Dobbs2018}. The spiral structure in the left panel of Figure \ref{fig:stacked_sfh_map} (0$-$79 Myr ago) shows multiple spiral arms, in agreement with the morphology that M33 shows in optical images. This correspondence suggests that the optical morphology is strongly shaped by the very low mass-to-light (M/L) ratio populations of the youngest stars. However, it important to note that this multi-arm spiral does seem to include the two spiral arms seen in the two-arm spiral structure shown by the morphology of the older stellar populations (older than 79 Myr, right panel of Figure \ref{fig:stacked_sfh_map}).

M33's lack of optically visible strong spiral arms has led to some debate in the literature about whether the spiral structure of M33 is a result of tidal interactions with M31 \citep[e.g.][]{Semczuk2018} or if this spiral structure arose without interaction \citep{vanderMarel2019}, as a result of gravitational instabilities in the stars and gas in M33 \citep{Dobbs2018}. Near infrared (NIR) observations in the J, H, K bands and with 2MASS have previously revealed a stronger two-arm spiral structure within the inner disk of M33 \citep{Regan&Vogel1994,Jarrett2003}.

The two-arm structure seen in the NIR agrees with our map of the SFH of M33 between 79 and 631 Myr ago. On the surface, this agreement may be surprising given the differing timescales of the two tracers. The NIR is typically presumed to be dominated by older red giant branch (RGB) and asymptotic giant branch (AGB) stars, far older than 500 Myr. However, there are two possible explanations. First, the NIR is in fact sensitive to red core Helium burning stars, which appear for ages $\lesssim$300 Myr \citep{Regan&Vogel1994,Melbourne2012}. Second, a strong long-lived spiral pattern in the stellar mass (as traced by the RGB stars that dominate the disk mass) can act as a gravitational perturbation that helps confine the younger stars, following global density wave theory \citep[e.g.,][]{lindblad1960,lin&shu1964}. Thus, while the gravitational-instability driven flocculent structure in the gas is imposed on stars at birth, this structure dissipates over $\sim$50 Myr as stars migrate from their birthplace and older stars are swept into the larger tidally-driven gravitational perturbation \citep[e.g.,][]{dobbs&pringle2010}.

A more detailed study of the structure of M33 with resolved stellar populations from the new PHATTER dataset is forthcoming, including a study of the age gradient across the spiral arms (A. Smercina et al., in preparation).

\subsection{Detection and Formation of M33 Bar}
There has been some debate in the literature about whether M33 hosts a central bar. A weak central bar has been observed via photometric measurements at infrared \citep[][]{Regan&Vogel1994,Block2004,Hernandez-Lopez2009} and blue wavelengths \citep{Elmegreen1992}, and via measurements of gas dynamics \citep{Corbelli2007}. However, \citet{Regan&Vogel1994} debated whether the weak central bar is actually an inner extension of the two spiral arms. Subsequent work studying gas dynamics in the center of M33 suggests that a small bar could explain discrepancies between models and the observed gas velocities in the inner (r$<$2kpc) disk \citep{Kam2015}.  Theoretically, \citet{Sellwood2019} and \citet{Semczuk2018} have both presented simulations that favor the formation of a bar in the inner 3 kpc of M33 within about 1 Gyr. For reference, the PHATTER survey extends $\sim$3.5 kpc from the center of M33, and $\sim$2 kpc along the semi-major axis.

\begin{figure*}
\centering
\includegraphics[width=0.95\textwidth]{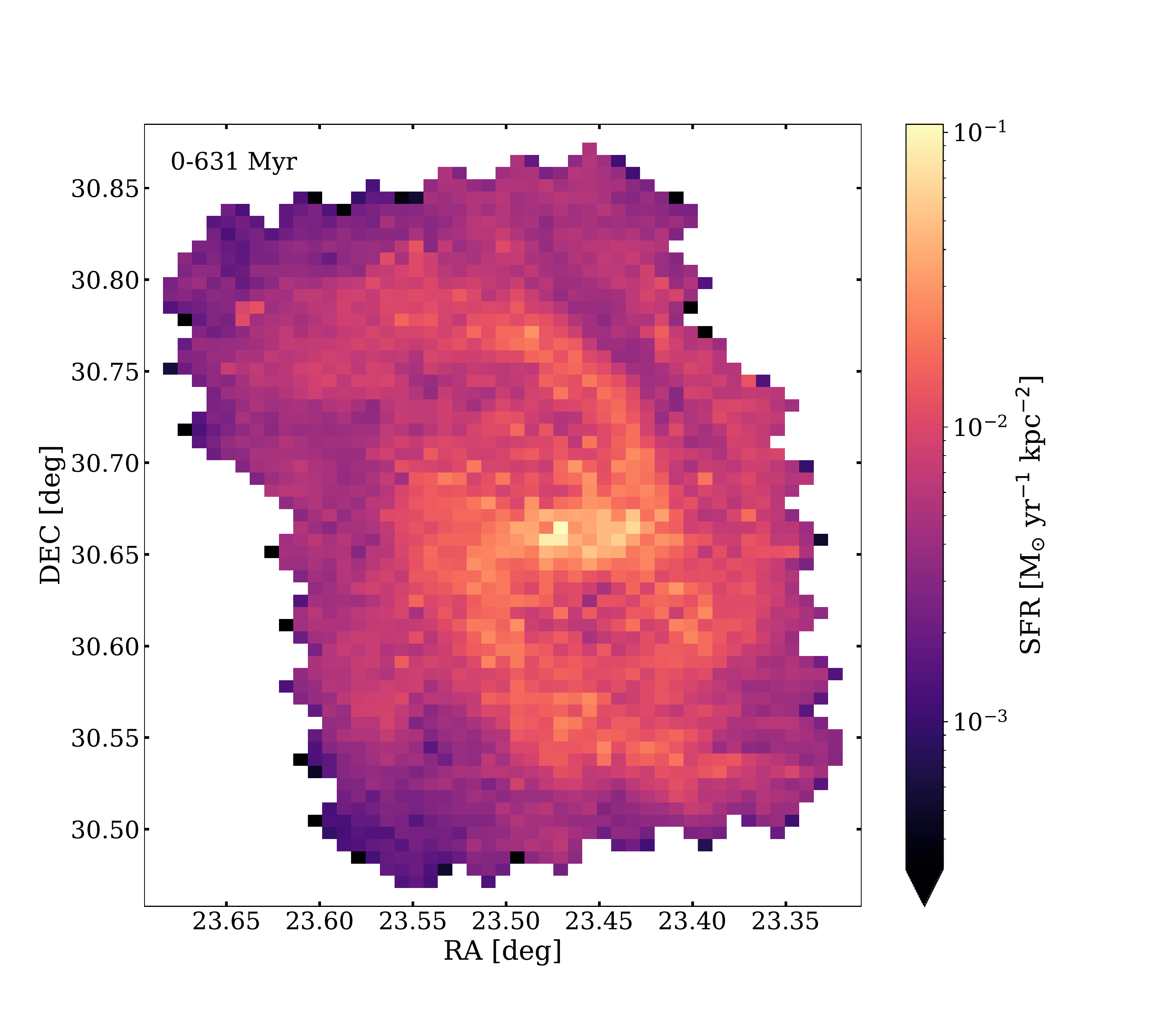}
\caption{The spatially resolved star formation rate of M33 between 0 and 631 Myr ago. This map is essentially a combination of the multiple SFH maps shown in Figure \ref{fig:sfr_log_bins_map} across temporal bins. The SFR value in each bin was calculated by summing the stellar mass formed in that spatial bin over all time bins between 0 and 631 Myr ago and then dividing by the total elapsed time. Each pixel represents a 100 pc by 100 pc region for which we derived the SFH using CMD-fitting.}
\label{fig:total_sfr_map}
\end{figure*}

While the young populations near the center of M33 appear to be confined to a relatively small and circular region, we observe a bar structure in the stellar distribution of stars older than roughly 79 Myr. This bar is also clearly visible in the integrated SFH map of star formation back to 630 Myr (Figure \ref{fig:total_sfr_map}). The maps of the integrated SFR in Figure \ref{fig:stacked_sfh_map} show no trace of the bar in populations younger than 79 Myr. The panel on the right, showing the SFR between 79 and 631 Myr ago, shows a clear bar extending out to $\sim$1 kpc. This suggests that the gas (which sets the spatial distribution of the youngest stars), does not trace the structure of the stellar bar, raising an interesting dynamical question about why the gas is able to resist formation of a bar, while the older stellar population is not.

More detailed analysis to fully characterize observations of the bar in M33 with the PHATTER data will be included in future work on the galaxy structure (A. Smercina et al., in preparation), but its appearance in the star formation history maps potentially gives us a timescale for stars to be perturbed into density waves and bar like orbits.

\subsection{Comparison of SFH results with other broadband measurements}
\begin{figure*}
\centering
\includegraphics[width=0.95\textwidth]{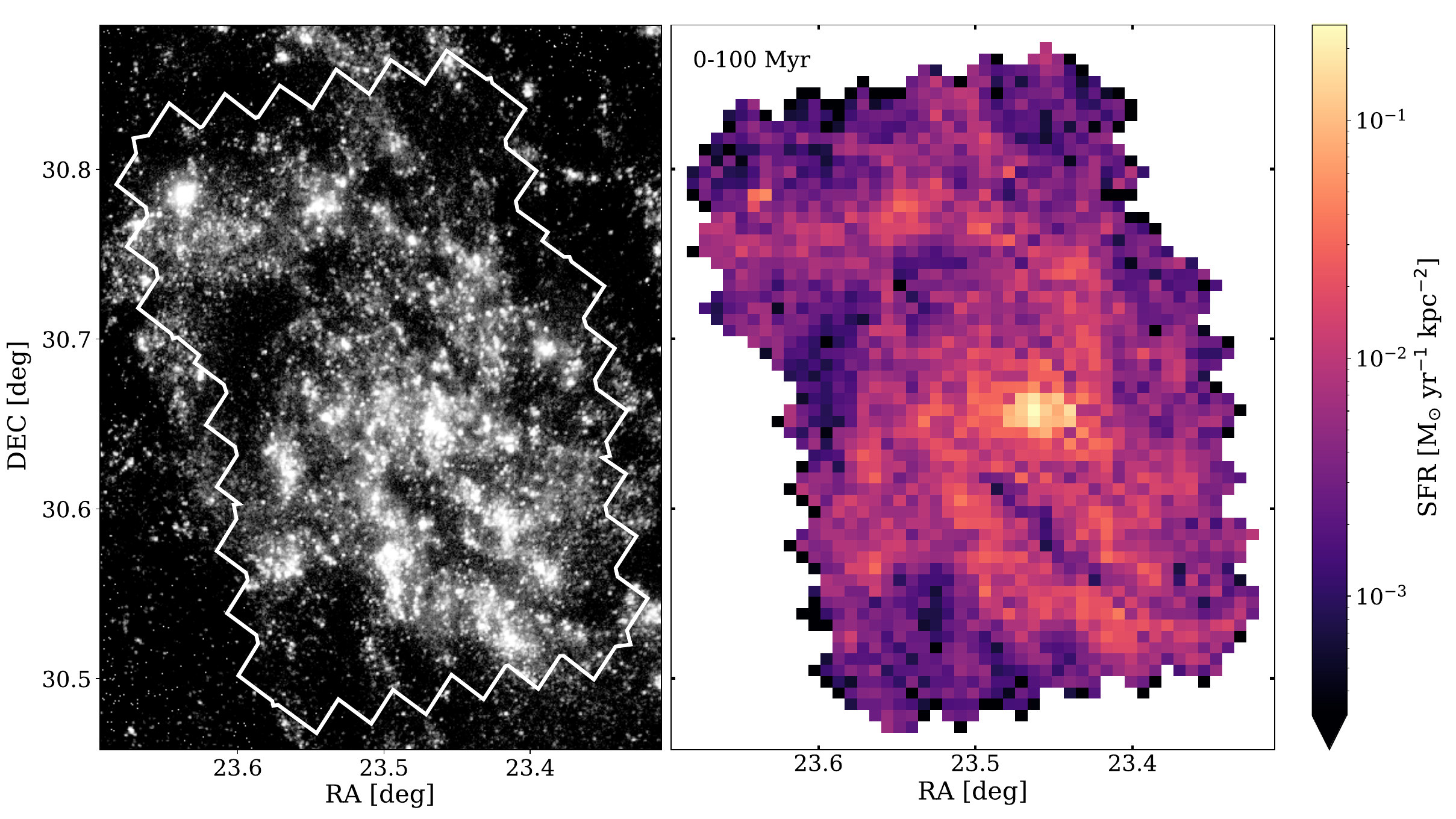}
\caption{The star formation rates we measure within the past 100 Myr trace the morphological structures visible in FUV images of M33. \textbf{left:} FUV image of M33, with the PHATTER survey outline overlaid. \textbf{right:} Star formation rate map of the M33 PHATTER survey within the past 100 Myr.}
\label{fig:100Myr_FUV_comparison}
\end{figure*}

We find that the morphology we observe in our 0$-$100 Myr SFH maps follows the morphology visible at FUV wavelengths. In Figure \ref{fig:100Myr_FUV_comparison}, we present a FUV image of M33 \citep{GildePaz2007} alongside a map of our SFR measurements in each spatial region over the past 100 Myr. Regions with high SFR within the past 100 Myr spatially coincide with high flux areas in the FUV image.

The PHATTER survey only covers the inner disk of M33, so to quantitatively compare our SFR measurements with published values, we have to correct our measured SFR for the star formation missed outside our survey area. For consistency with published work, we correct to an elliptical aperture within the D25 isophote, using the FUV D25 ellipse from \citet{GildePaz2007}. We then measure the total FUV flux within this D25 ellipse and determine the fraction of the flux contained within the PHATTER survey footprint. We find that the fraction of total flux within our survey footprint is $\sim$45\%, meaning we need to multiply our total SFR measurement by a factor of $\sim$2.2 to compare with global FUV SFR measurements. When we scale up our SFR by this factor, we get a total SFR of $\sim$0.74 M$_{\odot}$ yr$^{-1}$ for M33 within the last 100 Myr.

This value can be compared to the analysis of \citet{Verley2009}, who measured the global SFR of M33 using a combination of H$\alpha$ and FUV flux measurements, using IR luminosities to correct for dust extinction. Their extinction corrected value for the global SFR in M33 over the last 100 Myr is 0.45$\pm$0.10 M$_{\odot}$ yr$^{-1}$. This measurement is lower than our scaled-up global SFR by a factor of about 1.6 and does not agree within errors, suggesting that the integrated light measurements may be missing some star formation that the resolved stars are able to capture. This underestimation agrees with findings by \citet{Lewis2017}, who found that the 24 $\mu$m corrected FUV flux underestimated the SFR by a factor of 2.3$-$2.5 when compared to CMD-based SFHs in M31.

\begin{figure*}
\centering
\includegraphics[width=0.95\textwidth]{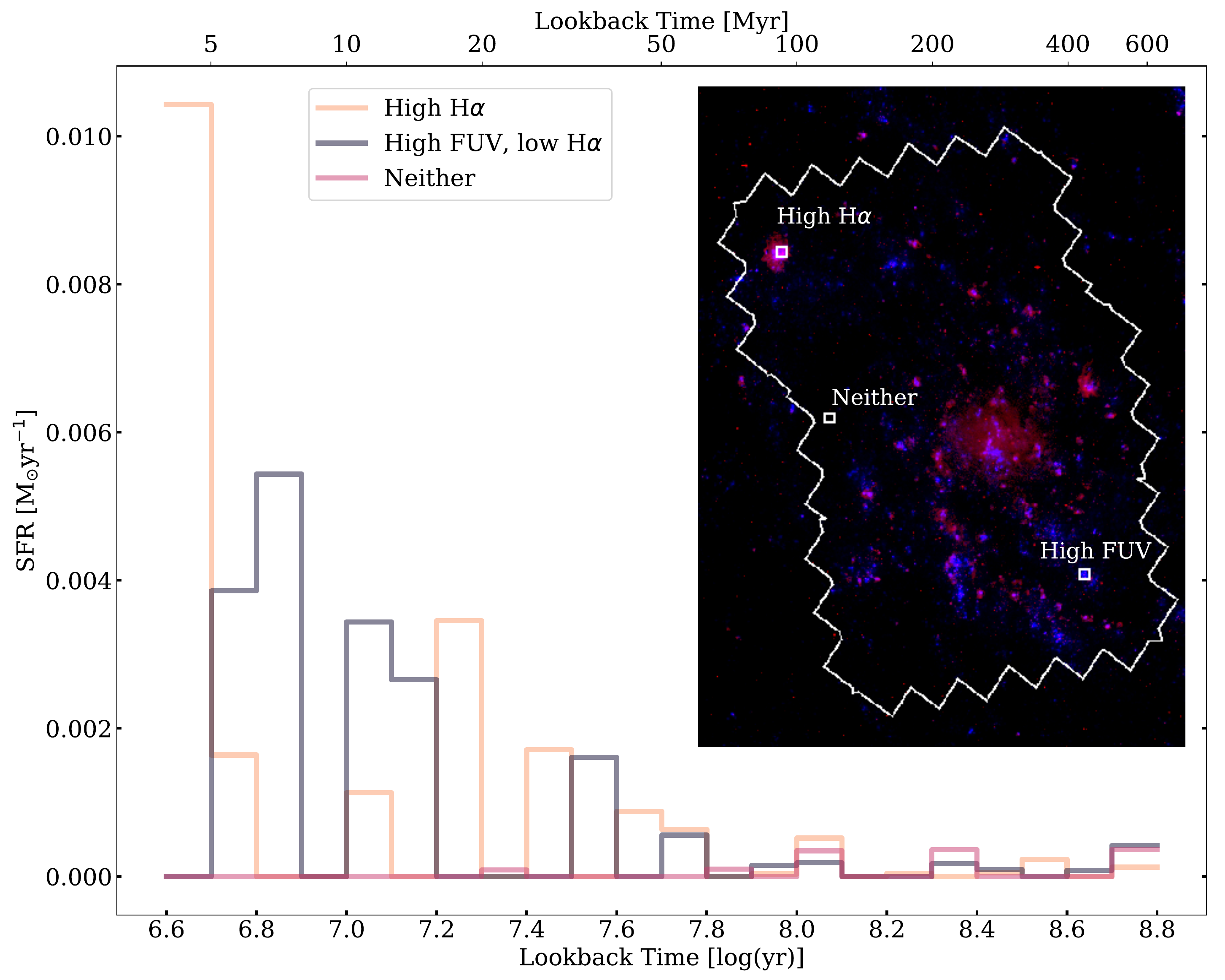}
\caption{Our best-fit measured SFHs reflect those we would expect based on SFR measurements in other bandpasses. In this figure we include a red-blue image of M33 in the inset with the PHATTER survey outlined in white. The blue band is a FUV image of M33 from GALEX \citep{GildePaz2007} and the red band is a H$\alpha$ image of M33 from the Local Group \citep{Massey2007}. We highlight three regions and plot the best-fit SFH for those regions in the main panel. We selected a region with high FUV flux, a region with high H$\alpha$ flux, and a region without high flux in either bandpass. The region without high flux in either FUV or H$\alpha$ shows little to no star formation in the last 100 Myr and low star formation rates beyond 100 Myr. The region selected for high H$\alpha$ flux shows significant star formation in the most recent time bin and high star formation rates back to about 100 Myr. The region selected for high FUV flux shows high star formation over the last 100 Myr but does not have measured star formation in the most recent time bin. These findings agree with our understanding of H$\alpha$ emission as a tracer of very recent (within the past 5 Myr), or instantaneous star formation \citep[e.g.;][]{Byler2017} and FUV emission as a tracer of star formation within the last 100$-$200 Myr as it directly traces emission from young, massive stars \citep[e.g.;][]{Kennicutt&Evans2012}. As we would expect, a region without significant H$\alpha$ or FUV emission does not show significant star formation within the past 100$-$200 Myr.}
\label{fig:halpha_fuv_comparison}
\end{figure*}

To check the correspondence between our SFHs and FUV emission more closely, we compared the SFH measured for some of our individual spatial regions that have high measured values of FUV and H$\alpha$ flux. In Figure \ref{fig:halpha_fuv_comparison} we present the SFH for three regions: one with relatively FUV flux, one with relatively high H$\alpha$ flux, and one with neither. We include a red-blue image where the red band is an H$\alpha$ image of M33 from the Local Group Galaxy Survey by \citet{Massey2007} and the blue band is a GALEX FUV image of M33 \citep{GildePaz2007}. 

The SFHs measured for these individual regions match our understanding of H$\alpha$ and FUV flux as tracers of recent star formation. We note that most regions that had high H$\alpha$ flux also had fairly significant FUV flux and visa versa. The region selected for high H$\alpha$ flux shows a peak in star formation within the last 5 Myr, but also has consistently high star formation over the past 100 Myr. The region selected for high FUV but low H$\alpha$ flux shows high star formation within the past 100 Myr, but does not show significant star formation within the past 5 Myr. The region that does not have significant FUV or H$\alpha$ flux shows essentially no star formation within the past 100 Myr and very low rates of star formation between 100 and 630 Myr, out to the oldest age considered in our analysis. This excellent correspondence suggests there is much more to be learned from the detailed comparisons of our SFH measurements and the UV emission (and other integrated light) characteristics of regions in M33, including comparisons with both local extinction and age. We plan to complete such comparisons in future work. 

\section{Conclusions}\label{sec:conclusions}
We have measured the SFH of M33 within the optical-only PHATTER survey footprint, which covers a de-projected area of $\sim$38 kpc$^{2}$ and extends to 3.5 kpc, 1$-$2 scale lengths, from the center of M33 \citep{Williams2021}. To measure the spatially resolved recent star formation history, we divided the survey area into 2005 analysis regions measuring 24$^{\prime \prime}$ (roughly 100 pc) on a side and measured the SFH for each region using color-magnitude diagram fitting. We summarize our main findings below:
\begin{enumerate}
    \item We measure the mean SFR over the last 100 Myr within the PHATTER survey area to be 0.32$\pm$0.02 M$_{\odot}$ yr$^{-1}$. When we scale up our measurement to account for the full D25 area of M33, we measure a mean SFR over the last 100 Myr of $\sim$0.74 M$_{\odot}$ yr$^{-1}$, which is higher than previous FUV 100 Myr SFR measurements.
    \item We measure the current SFR (within the past 10 Myr) within the PHATTER survey area of M33 to be 0.20$\pm$0.03 M$_{\odot}$ yr$^{-1}$.
    \item We observe evolution in the spiral arm structure of M33 over time. The most recent SFH maps of M33 show a flocculent multi-armed structure, while the maps older than $\sim79$ Myr ago show a more distinct two-armed structure.
    \item We observe a bar in the center of M33, which is observed in stellar populations older than $\sim79$ Myr. Younger stellar populations formed in more recent star formation episodes do not appear to follow the bar structure.
\end{enumerate}

\begin{acknowledgments}
We thank Leo Girardi for his thoughtful comments on this paper. Support for this work was provided by NASA through grant \#GO-14610 from the Space Telescope Science Institute, which is operated by AURA, Inc., under NASA contract NAS 5-26555. Margaret Lazzarini was supported by an NSF Astronomy and Astrophysics Postdoctoral Fellowship under award AST-2102721 during a portion of this work.
\end{acknowledgments}
\begin{deluxetable*}{cll}
\tablecaption{Column Information for Table \ref{table:big_preview} \label{table:col_info}}
\tablehead{
\colhead{Column Number} &
\colhead{Column Name} &
\colhead{Column Description}
}
\startdata
1& Region ID&  Reference ID for each spatial region\\ 
2& R.A. 1&  R.A. of North East corner of spatial region [deg]\\ 
3& Dec. 1&  Dec. of North East corner of spatial region [deg]\\ 
4& R.A. 2&  R.A. of North West corner of spatial region [deg]\\ 
5& Dec. 2&  Dec. of North West corner of spatial region [deg]\\ 
6& R.A. 3&  R.A. of South West corner of spatial region [deg]\\ 
7& Dec. 3&  Dec. of South West corner of spatial region [deg]\\ 
8& R.A. 4&  R.A. of South East corner of spatial region [deg]\\ 
9& Dec. 4&  Dec. of South East corner of spatial region [deg]\\ 
10& SFR 6.6-6.7&  SFR between log(years)=6.6 and log(years)=6.7 [$\times10^{-5}$ M$_{\odot}$ yr$^{-1}$]\\ 
11& SFR 6.7-6.8& SFR between log(years)=6.7 and log(years)=6.8 [$\times10^{-5}$ M$_{\odot}$ yr$^{-1}$] \\ 
12& SFR 6.8-6.9& SFR between log(years)=6.8 and log(years)=6.9 [$\times10^{-5}$ M$_{\odot}$ yr$^{-1}$] \\ 
13& SFR 6.9-7.0& SFR between log(years)=6.9 and log(years)=7.0 [$\times10^{-5}$ M$_{\odot}$ yr$^{-1}$] \\ 
14& SFR 7.0-7.1&  SFR between log(years)=7.0 and log(years)=7.1 [$\times10^{-5}$ M$_{\odot}$ yr$^{-1}$]\\ 
15& SFR 7.1-7.2&  SFR between log(years)=7.1 and log(years)=7.2 [$\times10^{-5}$ M$_{\odot}$ yr$^{-1}$]\\ 
16& SFR 7.2-7.3&  SFR between log(years)=7.2 and log(years)=7.3 [$\times10^{-5}$ M$_{\odot}$ yr$^{-1}$]\\ 
17& SFR 7.3-7.4&  SFR between log(years)=7.3 and log(years)=7.4 [$\times10^{-5}$ M$_{\odot}$ yr$^{-1}$]\\ 
18& SFR 7.4-7.5&  SFR between log(years)=7.4 and log(years)=7.5 [$\times10^{-5}$ M$_{\odot}$ yr$^{-1}$]\\ 
19& SFR 7.5-7.6&  SFR between log(years)=7.5 and log(years)=7.6 [$\times10^{-5}$ M$_{\odot}$ yr$^{-1}$]\\ 
20& SFR 7.6-7.7&  SFR between log(years)=7.6 and log(years)=7.7 [$\times10^{-5}$ M$_{\odot}$ yr$^{-1}$]\\ 
21& SFR 7.7-7.8&  SFR between log(years)=7.7 and log(years)=7.8 [$\times10^{-5}$ M$_{\odot}$ yr$^{-1}$]\\ 
22& SFR 7.8-7.9&  SFR between log(years)=7.8 and log(years)=7.9 [$\times10^{-5}$ M$_{\odot}$ yr$^{-1}$]\\ 
23& SFR 7.9-8.0&  SFR between log(years)=7.9 and log(years)=8.0 [$\times10^{-5}$ M$_{\odot}$ yr$^{-1}$]\\ 
24& SFR 8.0-8.1& SFR between log(years)=8.0 and log(years)=8.1 [$\times10^{-5}$ M$_{\odot}$ yr$^{-1}$] \\ 
25& SFR 8.1-8.2&  SFR between log(years)=8.1 and log(years)=8.2 [$\times10^{-5}$ M$_{\odot}$ yr$^{-1}$]\\ 
26& SFR 8.2-8.3&  SFR between log(years)=8.2 and log(years)=8.3 [$\times10^{-5}$ M$_{\odot}$ yr$^{-1}$]\\ 
27& SFR 8.3-8.4&  SFR between log(years)=8.3 and log(years)=8.4 [$\times10^{-5}$ M$_{\odot}$ yr$^{-1}$]\\ 
28& SFR 8.4-8.5&  SFR between log(years)=8.4 and log(years)=8.5 [$\times10^{-5}$ M$_{\odot}$ yr$^{-1}$]\\ 
29& SFR 8.5-8.6&  SFR between log(years)=8.5 and log(years)=8.6 [$\times10^{-5}$ M$_{\odot}$ yr$^{-1}$]\\ 
30& SFR 8.6-8.7&  SFR between log(years)=8.6 and log(years)=8.7 [$\times10^{-5}$ M$_{\odot}$ yr$^{-1}$]\\ 
31& SFR 8.7-8.8&  SFR between log(years)=8.7 and log(years)=8.8 [$\times10^{-5}$ M$_{\odot}$ yr$^{-1}$]\\ 
32& $A_{V}$ & Best-fit foreground extinction [mag] \\
33& $dA_{V}$ & Best-fit differential extinction [mag] \\
34& Area & Area covered by optical-only catalog [arcsec$^{2}$] \\
35& SFR 0-10 Myr & SFR between 0 and 10 Myr [$\times10^{-5}$ M$_{\odot}$ yr$^{-1}$] \\
36& SFR 0-100 Myr & SFR between 0 and 100 Myr [$\times10^{-5}$ M$_{\odot}$ yr$^{-1}$] \\
37& SFR 0-500 Myr & SFR between 0 and 500 Myr [$\times10^{-5}$ M$_{\odot}$ yr$^{-1}$] \\
38& SFR 0-6300 Myr & SFR between 0 and 630 Myr [$\times10^{-5}$ M$_{\odot}$ yr$^{-1}$]
\enddata 
\end{deluxetable*}
\begin{splitdeluxetable*}{cccccccccccccBccccccccccccBccccccccccccc}
\tabletypesize{\scriptsize}
\tablecaption{SFH for Individual Spatial Regions in M33 \label{table:big_preview}}
\rotate
\tablehead{
\colhead{Region ID} &
\colhead{R.A. 1} &
\colhead{Dec. 1} &
\colhead{R.A. 2} &
\colhead{Dec. 2} &
\colhead{R.A. 3} &
\colhead{Dec. 3} &
\colhead{R.A. 4} &
\colhead{Dec. 4} &
\colhead{SFR 6.6-6.7} &
\colhead{SFR 6.7-6.8} &
\colhead{SFR 6.8-6.9} &
\colhead{SFR 6.9-7.0} &
\colhead{SFR 7.0-7.1} &
\colhead{SFR 7.1-7.2} &
\colhead{SFR 7.2-7.3} &
\colhead{SFR 7.3-7.4} &
\colhead{SFR 7.4-7.5} &
\colhead{SFR 7.5-7.6} &
\colhead{SFR 7.6-7.7} &
\colhead{SFR 7.7-7.8} &
\colhead{SFR 7.8-7.9} &
\colhead{SFR 7.9-8.0} &
\colhead{SFR 8.0-8.1} &
\colhead{SFR 8.1-8.2} &
\colhead{SFR 8.2-8.3} &
\colhead{SFR 8.3-8.4} &
\colhead{SFR 8.4-8.5} &
\colhead{SFR 8.5-8.6} &
\colhead{SFR 8.6-8.7} &
\colhead{SFR 8.7-8.8} &
\colhead{$A_{V}$} &
\colhead{$dA_{V}$} &
\colhead{Area} &
\colhead{SFR 0-10 Myr} &
\colhead{SFR 0-100 Myr} &
\colhead{SFR 0-500 Myr} &
\colhead{SFR 0-630 Myr} \\
(1) & (2) & (3) & (4) & (5) & (6) & (7) & (8) & (9) & (10) & (11) & (12) & (13) & (14) & (15) & (16) & (17) & (18) & (19) & (20) & (21) & (22) & (23) & (24) & (25) & (26) & (27) & (28) & (29) & (30) & (31) & (32) & (33) & (34) & (35) & (36) & (37) & (38)}
\startdata
\vspace{-0.25cm}\\
1 & 23.513372 & 30.474611 & 23.505617 & 30.474611 & 23.505617 & 30.467941 & 23.513372 & 30.467941 & $0.0^{+3.2}_{-0.0}$ & $0.0^{+2.6}_{-0.0}$ & $0.0^{+2.0}_{-0.0}$ & $0.0^{+1.7}_{-0.0}$ & $0.0^{+1.4}_{-0.0}$ & $0.0^{+1.1}_{-0.0}$ & $0.0^{+0.9}_{-0.0}$ & $0.0^{+0.8}_{-0.0}$ & $0.0^{+0.6}_{-0.0}$ & $0.0^{+0.5}_{-0.0}$ & $0.0^{+0.4}_{-0.0}$ & $0.0^{+0.3}_{-0.0}$ & $0.0^{+0.3}_{-0.0}$ & $0.0^{+0.2}_{-0.0}$ & $0.0^{+0.2}_{-0.0}$ & $0.0^{+0.2}_{-0.0}$ & $0.0^{+0.2}_{-0.0}$ & $0.09^{+0.1}_{-0.1}$ & $0.0^{+0.2}_{-0.0}$ & $0.39^{+0.0}_{-0.3}$ & $0.0^{+0.2}_{-0.0}$ & $0.19^{+0.1}_{-0.2}$ & 0.16 & 0.05 & 20 & $0.0^{+1.4}_{-0.0}$ & $0.0^{+0.5}_{-0.0}$ & $0.07^{+0.2}_{-0.1}$ & $0.1^{+0.2}_{-0.1}$\\
2 & 23.544392 & 30.474611 & 23.536637 & 30.474611 & 23.536637 & 30.467941 & 23.544392 & 30.467941 & $0.0^{+10.5}_{-0.0}$ & $0.0^{+9.0}_{-0.0}$ & $0.0^{+7.3}_{-0.0}$ & $0.0^{+6.1}_{-0.0}$ & $0.0^{+5.3}_{-0.0}$ & $0.0^{+4.6}_{-0.0}$ & $0.0^{+3.9}_{-0.0}$ & $0.0^{+3.2}_{-0.0}$ & $0.0^{+2.8}_{-0.0}$ & $0.0^{+2.6}_{-0.0}$ & $6.68^{+0.0}_{-5.5}$ & $0.0^{+2.0}_{-0.0}$ & $0.0^{+1.7}_{-0.0}$ & $0.0^{+1.7}_{-0.0}$ & $0.0^{+1.9}_{-0.0}$ & $4.76^{+0.0}_{-3.8}$ & $0.0^{+1.2}_{-0.0}$ & $0.0^{+0.8}_{-0.0}$ & $0.0^{+0.6}_{-0.0}$ & $0.0^{+0.6}_{-0.0}$ & $0.0^{+0.5}_{-0.0}$ & $0.43^{+0.3}_{-0.4}$ & 0.31 & 0.1 & 104 & $0.0^{+4.7}_{-0.0}$ & $0.69^{+2.4}_{-0.6}$ & $0.45^{+1.0}_{-0.4}$ & $0.44^{+0.9}_{-0.4}$\\
3 & 23.552147 & 30.474611 & 23.544392 & 30.474611 & 23.544392 & 30.467941 & 23.552147 & 30.467941 & $0.0^{+14.9}_{-0.0}$ & $0.0^{+13.0}_{-0.0}$ & $0.0^{+10.1}_{-0.0}$ & $6.89^{+2.6}_{-6.9}$ & $0.0^{+8.0}_{-0.0}$ & $0.0^{+6.7}_{-0.0}$ & $0.0^{+5.7}_{-0.0}$ & $0.0^{+5.4}_{-0.0}$ & $0.0^{+4.5}_{-0.0}$ & $0.0^{+4.4}_{-0.0}$ & $0.0^{+4.8}_{-0.0}$ & $8.8^{+4.2}_{-7.4}$ & $23.21^{+0.0}_{-13.5}$ & $0.0^{+6.5}_{-0.0}$ & $0.99^{+3.6}_{-1.0}$ & $0.0^{+3.3}_{-0.0}$ & $0.0^{+3.2}_{-0.0}$ & $8.32^{+0.8}_{-5.1}$ & $0.0^{+3.1}_{-0.0}$ & $2.16^{+1.2}_{-2.1}$ & $0.65^{+2.3}_{-0.6}$ & $4.19^{+0.6}_{-2.6}$ & 0.16 & 0.2 & 448 & $1.42^{+5.4}_{-1.4}$ & $5.08^{+4.5}_{-3.3}$ & $2.41^{+2.7}_{-1.7}$ & $2.77^{+2.3}_{-1.9}$\\
4 & 23.559902 & 30.474611 & 23.552147 & 30.474611 & 23.552147 & 30.467941 & 23.559902 & 30.467941 & $0.0^{+11.5}_{-0.0}$ & $0.0^{+9.6}_{-0.0}$ & $0.0^{+7.9}_{-0.0}$ & $0.0^{+6.8}_{-0.0}$ & $0.0^{+6.1}_{-0.0}$ & $0.0^{+5.1}_{-0.0}$ & $0.0^{+4.4}_{-0.0}$ & $0.0^{+7.0}_{-0.0}$ & $17.47^{+0.2}_{-12.1}$ & $0.0^{+6.1}_{-0.0}$ & $5.47^{+0.0}_{-4.8}$ & $0.0^{+2.6}_{-0.0}$ & $0.0^{+1.5}_{-0.0}$ & $0.0^{+1.0}_{-0.0}$ & $0.0^{+0.8}_{-0.0}$ & $0.0^{+0.6}_{-0.0}$ & $0.14^{+0.4}_{-0.1}$ & $0.0^{+0.6}_{-0.0}$ & $0.96^{+0.0}_{-0.9}$ & $0.0^{+0.4}_{-0.0}$ & $0.0^{+0.4}_{-0.0}$ & $0.13^{+0.3}_{-0.1}$ & 0.36 & 0.0 & 82 & $0.0^{+5.1}_{-0.0}$ & $1.7^{+2.7}_{-1.3}$ & $0.48^{+0.9}_{-0.4}$ & $0.41^{+0.7}_{-0.3}$\\
5 & 23.474598 & 30.481281 & 23.466843 & 30.481281 & 23.466843 & 30.474611 & 23.474598 & 30.474611 & $0.0^{+5.5}_{-0.0}$ & $0.0^{+4.6}_{-0.0}$ & $0.0^{+3.8}_{-0.0}$ & $0.0^{+3.0}_{-0.0}$ & $0.0^{+2.5}_{-0.0}$ & $0.0^{+2.0}_{-0.0}$ & $0.0^{+1.7}_{-0.0}$ & $0.0^{+1.4}_{-0.0}$ & $0.0^{+1.3}_{-0.0}$ & $0.0^{+1.0}_{-0.0}$ & $0.0^{+0.9}_{-0.0}$ & $0.0^{+0.8}_{-0.0}$ & $0.0^{+0.7}_{-0.0}$ & $0.0^{+0.7}_{-0.0}$ & $0.0^{+0.7}_{-0.0}$ & $0.0^{+0.8}_{-0.0}$ & $1.87^{+0.0}_{-1.6}$ & $0.14^{+0.4}_{-0.1}$ & $0.0^{+0.6}_{-0.0}$ & $0.0^{+0.9}_{-0.0}$ & $2.21^{+0.1}_{-1.3}$ & $0.0^{+1.0}_{-0.0}$ & 0.11 & 0.3 & 60 & $0.0^{+2.4}_{-0.0}$ & $0.0^{+1.1}_{-0.0}$ & $0.62^{+0.6}_{-0.4}$ & $0.49^{+0.7}_{-0.3}$\\
6 & 23.505617 & 30.481281 & 23.497862 & 30.481281 & 23.497862 & 30.474611 & 23.505617 & 30.474611 & $0.0^{+9.5}_{-0.0}$ & $0.0^{+7.4}_{-0.0}$ & $0.0^{+5.9}_{-0.0}$ & $0.0^{+5.4}_{-0.0}$ & $0.0^{+4.3}_{-0.0}$ & $4.56^{+0.0}_{-4.3}$ & $0.0^{+3.1}_{-0.0}$ & $0.0^{+2.4}_{-0.0}$ & $0.0^{+2.0}_{-0.0}$ & $0.0^{+1.7}_{-0.0}$ & $0.0^{+2.0}_{-0.0}$ & $2.83^{+0.0}_{-2.7}$ & $2.34^{+0.0}_{-2.3}$ & $0.0^{+1.5}_{-0.0}$ & $0.0^{+0.9}_{-0.0}$ & $0.0^{+0.7}_{-0.0}$ & $0.43^{+0.2}_{-0.4}$ & $0.0^{+0.6}_{-0.0}$ & $0.0^{+0.5}_{-0.0}$ & $0.99^{+0.0}_{-1.0}$ & $0.0^{+1.1}_{-0.0}$ & $1.83^{+0.6}_{-1.4}$ & 0.51 & 0.0 & 89 & $0.0^{+4.0}_{-0.0}$ & $0.9^{+1.5}_{-0.9}$ & $0.38^{+0.8}_{-0.4}$ & $0.68^{+0.7}_{-0.6}$\\
\enddata 
\tablecomments{We only include the SFH for the first 6 spatial regions in the print version of this paper. This table is available in its entirety in machine-readable form in the online version of this paper. See Table \ref{table:col_info} for a description of each column.}
\end{splitdeluxetable*}
\pagebreak
\begin{deluxetable}{ccc}
\tablecaption{Total PHATTER SFH \label{table:total_sfh}}
\tablehead{
\colhead{Age Range } &
\colhead{Age Range} &  
\colhead{Star Formation Rate} \\
\colhead{[log(years)]} &
\colhead{[Myr]} &
\colhead{[M$_{\odot}$ yr$^{-1}$]}
}
\startdata
6.6$-$6.7 & 4$-$5 & 0.11 $^{+ 0.04 }_{- 0.04 }$ \\
6.7$-$6.8 & 5$-$6 & 0.38 $^{+ 0.04 }_{- 0.03 }$ \\
6.8$-$6.9 & 6$-$8 & 0.27 $^{+ 0.03 }_{- 0.02 }$ \\
6.9$-$7.0 & 8$-$10 & 0.25 $^{+ 0.02 }_{- 0.02 }$ \\
7.0$-$7.1 & 10$-$13 & 0.32 $^{+ 0.02 }_{- 0.02 }$ \\
7.1$-$7.2 & 13$-$16 & 0.31 $^{+ 0.02 }_{- 0.02 }$ \\
7.2$-$7.3 & 16$-$20 & 0.31 $^{+ 0.02 }_{- 0.02 }$ \\
7.3$-$7.4 & 20$-$25 & 0.27 $^{+ 0.02 }_{- 0.02 }$ \\
7.4$-$7.5 & 25$-$32 & 0.30 $^{+ 0.01 }_{- 0.02 }$ \\
7.5$-$7.6 & 32$-$40 & 0.30 $^{+ 0.02 }_{- 0.02 }$ \\
7.6$-$7.7 & 40$-$50 & 0.33 $^{+ 0.02 }_{- 0.02 }$ \\
7.7$-$7.8 & 50$-$63 & 0.33 $^{+ 0.02 }_{- 0.01 }$ \\
7.8$-$7.9 & 63$-$79 & 0.37 $^{+ 0.02 }_{- 0.01 }$ \\
7.9$-$8.0 & 79$-$100 & 0.37 $^{+ 0.01 }_{- 0.02 }$ \\
8.0$-$8.1 & 100$-$126 & 0.30 $^{+ 0.02 }_{- 0.01 }$ \\
8.1$-$8.2 & 126$-$158 & 0.29 $^{+ 0.01 }_{- 0.01 }$ \\
8.2$-$8.3 & 158$-$200 & 0.29 $^{+ 0.01 }_{- 0.01 }$ \\
8.3$-$8.4 & 200$-$251 & 0.28 $^{+ 0.01 }_{- 0.01 }$ \\
8.4$-$8.5 & 251$-$316 & 0.26 $^{+ 0.01 }_{- 0.01 }$ \\
8.5$-$8.6 & 316$-$398 & 0.32 $^{+ 0.01 }_{- 0.01 }$ \\
8.6$-$8.7 & 398$-$501 & 0.40 $^{+ 0.01 }_{- 0.01 }$ \\
8.7$-$8.8 & 501$-$631 & 0.40 $^{+ 0.01 }_{- 0.01 }$ \\
\enddata
\tablecomments{The spatially-integrated SFH of the PHATTER survey in M33. The first column lists the start of each time bin in log(years) and the second column lists the end of each time bin in log(years). The third column lists the SFR for that time bin with uncertainties. The listed uncertainties incorporates both random and dust uncertainties. For more information on how these uncertainties were derived, see Section \ref{sec:uncertainties}.}
\end{deluxetable}
\begin{deluxetable*}{ccc}
\tablecaption{Integrated SFR Over Various Timescales \label{table:sfr_timescales}}
\tablehead{
\colhead{Time Range} &
\colhead{SFR [M$_{\odot}$ yr$^{-1}$]}  &
\colhead{SFR [M$_{\odot}$ yr$^{-1}$ kpc$^{-2}$]}
}
\startdata
0-10 Myr &  0.20 +0.04/-0.03 & 5.2$\pm$0.9$\times$10$^{-3}$ \\
0-100 Myr & 0.32$\pm$0.02 & 8.4$\pm$0.5$\times$10$^{-3}$\\
0-500 Myr & 0.32$\pm$0.01 & 8.4$\pm$0.3$\times$10$^{-3}$ \\
0-630 Myr & 0.34$\pm$0.01 & 8.8$\pm$0.3$\times$10$^{-3}$ 
\enddata
\tablecomments{Table listing the spatially and temporally integrated SFR across the PHATTER footprint over various timescales.}
\end{deluxetable*} 
\clearpage
\appendix
\section{Comparing Model CMDs with Observed CMDs}\label{sec:cmd_residuals}
\begin{figure*}
\centering
\includegraphics[width=0.95\textwidth]{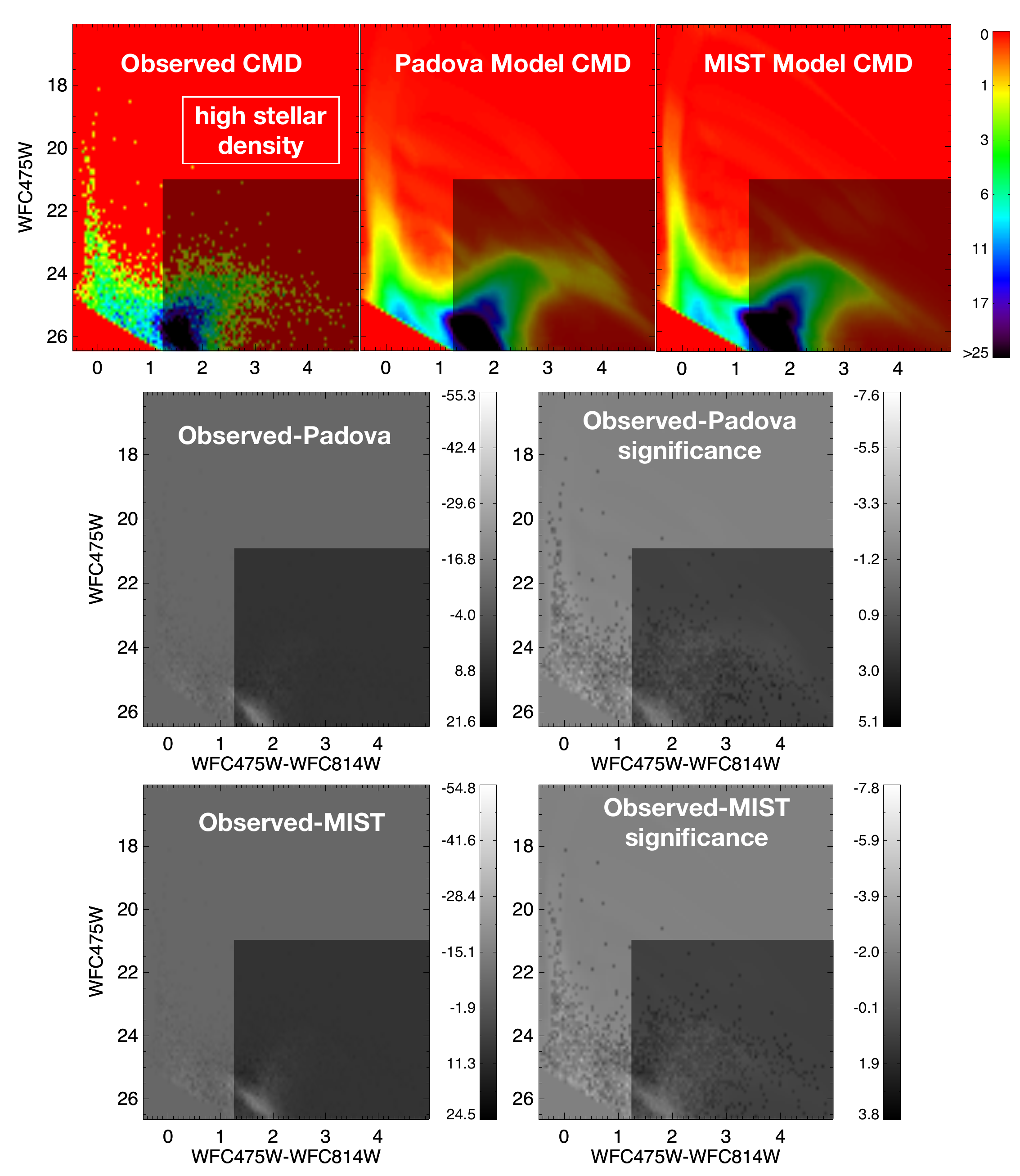}
\caption{Comparison between the observed and model CMDs generated with the Padova and MIST stellar models for a high stellar density region, 843. The dark shaded area on each panel shows the portion of the CMD that is excluded in our fits. \textbf{Top:} The observed, Padova model, and MIST model CMDs for a high stellar density region, 843. The color-map indicates the number of stars in each region of the CMD. \textbf{Middle:} The difference between the observed and Padova model CMD (left) and the residual significance (right), which is the observed CMD minus the modeled CMD, weighted by the variance. \textbf{Bottom:} The difference between the observed and MIST model CMD (left) and the residual significance (right). We find no systematic residuals in the area of the CMD included in our fits, indicating that both sets of stellar models are a good fit to the data.}
\label{fig:hi_density_CMD_residuals}
\end{figure*}

\begin{figure*}
\centering
\includegraphics[width=0.95\textwidth]{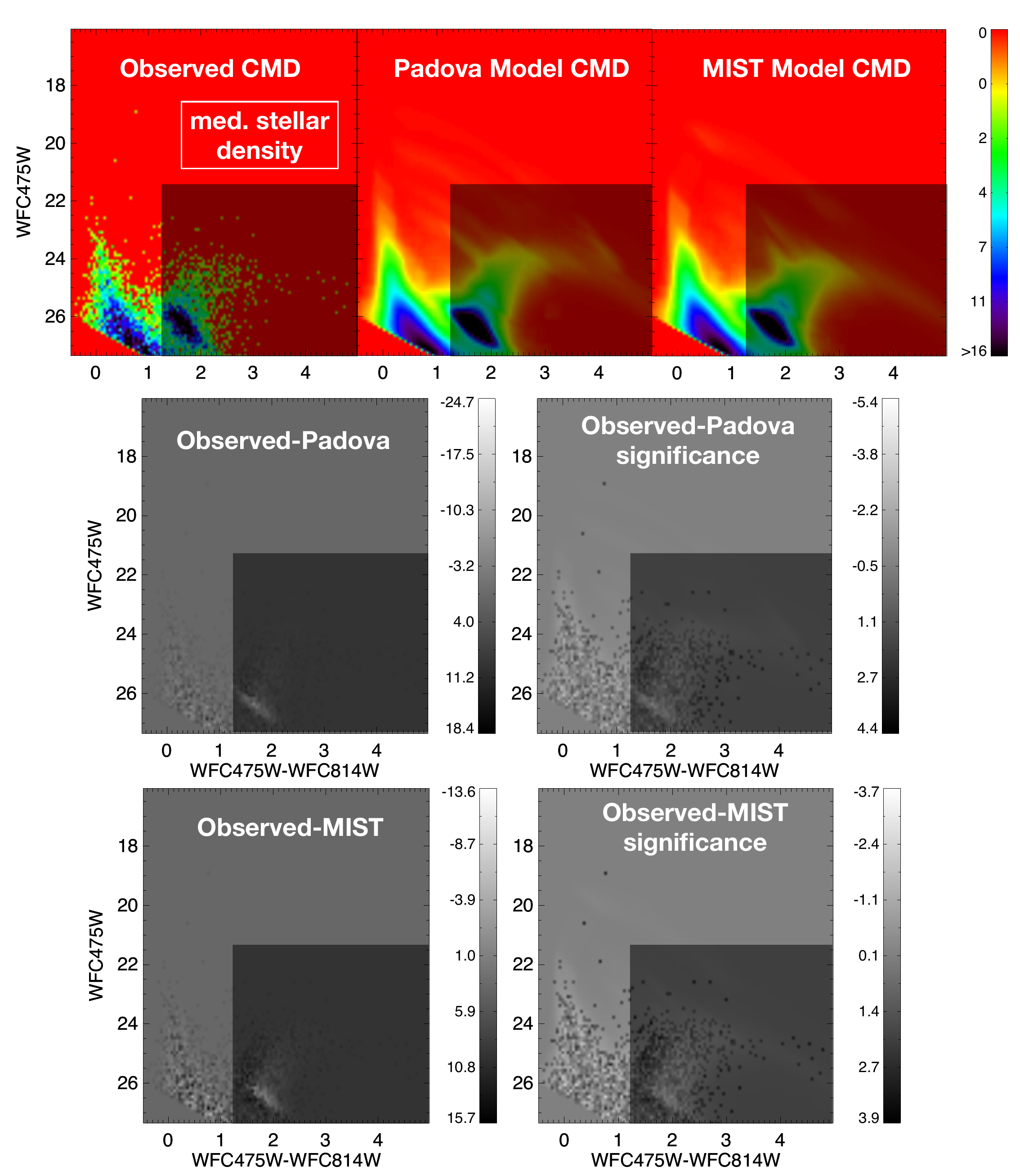}
\caption{Comparison between the observed and model CMDs generated with the Padova and MIST stellar models for a medium stellar density region, 1202. The dark shaded area on each panel shows the portion of the CMD that is excluded in our fits. \textbf{Top:} The observed, Padova model, and MIST model CMDs for a medium stellar density region, 1202. The color-map indicates the number of stars in each region of the CMD. \textbf{Middle:} The difference between the observed and Padova model CMD (left) and the residual significance (right), which is the observed CMD minus the modeled CMD, weighted by the variance. \textbf{Bottom:} The difference between the observed and MIST model CMD (left) and the residual significance (right). We find no systematic residuals in the area of the CMD included in our fits, indicating that both sets of stellar models are a good fit to the data.}
\label{fig:med_density_CMD_residuals}
\end{figure*}

\begin{figure*}
\centering
\includegraphics[width=0.95\textwidth]{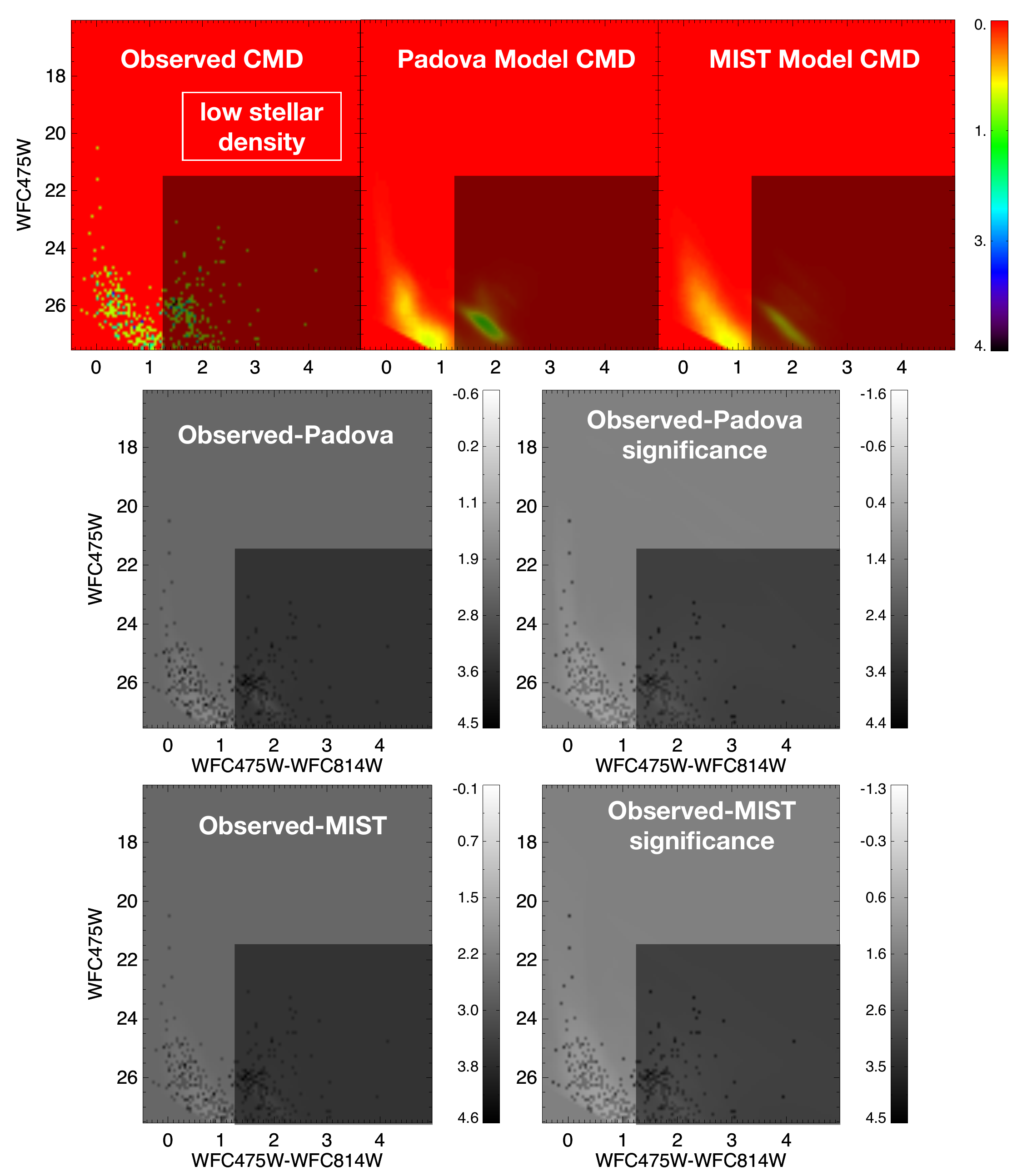}
\caption{Comparison between the observed and model CMDs generated with the Padova and MIST stellar models for a low stellar density region, 1404. The dark shaded area on each panel shows the portion of the CMD that is excluded in our fits. \textbf{Top:} The observed, Padova model, and MIST model CMDs for a low stellar density region, 1404. The color-map indicates the number of stars in each region of the CMD. \textbf{Middle:} The difference between the observed and Padova model CMD (left) and the residual significance (right), which is the observed CMD minus the modeled CMD, weighted by the variance. \textbf{Bottom:} The difference between the observed and MIST model CMD (left) and the residual significance (right). We find no systematic residuals in the area of the CMD included in our fits, indicating that both sets of stellar models are a good fit to the data}
\label{fig:lo_density_CMD_residuals}
\end{figure*}

In our fitting process, we generate model CMDs which can be compared against the observed CMDs to assess the quality of our fits. For each region being fit, we can then divide the CMD into bins and compare the number of observed stars in each bin to the number of model CMD stars in each bin. We can then find the difference between these two values and calculate the significance of this difference. The significance is calculated using a Poisson maximum likelihood statistic, based on the Poisson equivalent of $\chi^{2}$ \citep{Dolphin2002}.
To demonstrate how well both the Padova and MIST models perform in creating model CMDs that fit the data, we have included figures (\ref{fig:hi_density_CMD_residuals}, \ref{fig:med_density_CMD_residuals}, and \ref{fig:lo_density_CMD_residuals}) in this appendix showing the observed and model CMDs for three regions with varying stellar density: region 843 (high density), region 1202 (medium density), and region 1404 (low density). These three regions all have approximately the median values of A$_{V}$ and dA$_{V}$ in our sample.
Each of Figures \ref{fig:hi_density_CMD_residuals} through \ref{fig:lo_density_CMD_residuals} includes seven panels. The shaded region in each panel shows the region of the CMD that is excluded in our SFH fits. The upper left panel shows the observed CMD, with the color showing the number of stars in each bin of the CMD. The upper middle panel shows the Padova model CMD, with the color showing the number of stars in each bin of the CMD. The upper right panel shows the MIST model CMD, with the color showing the number of stars in each bin of the CMD. The middle left panel shows the difference between the observed and Padova model CMD and the middle right panel shows the significance of the difference, which is described above. The lower left panel shows the difference between the observed and MIST model CMD and the lower right panel shows the significance of the difference.
We can analyze the significance of the difference between the model and observed CMDs to look for 1. differences between how well the MIST and Padova model CMDs fit the data and 2. trends that could suggest systematic bias in our SFH measurements. In Figures \ref{fig:hi_density_CMD_residuals}, \ref{fig:med_density_CMD_residuals}, and \ref{fig:lo_density_CMD_residuals} there is not a clear difference in the significance of the difference between the model and observed CMDs for the MIST and Padova models, which we can see when comparing the middle right and lower right panels of these figures. We also do not see a strong trend in the significance of the difference between the model and observed CMDs in the areas of the CMD being fit to recover the SFH. Specifically, the difference between the model and observed CMDs looks fairly even across the CMD, rather than demonstrating that our models under or over produce stars in different parts of the CMD that are outside of the shaded region excluded in our fits.
The significance of the difference between the model CMDs and real CMDs is used to calculate the goodness of the fit, which is based on the Poisson equivalent of $\chi^{2}$. To get a holistic view of how well the Padova and MIST models represent the whole data set, we compared the fit values for each of the 2006 regions in our SFH maps. We found that the distribution of fit values across all regions was very similar for the Padova and MIST models. The Padova mean fit value across all regions agrees with the MIST mean fit value across all regions within $<$1\%. This suggests that while there are some differences at young ages ($<$20 Myr) between the Padova and MIST models due to theoretical uncertainties, both sets of stellar models are similarly successful at matching the data.
\section{SFH Recovery Tests}\label{sec:sfh_recovery_tests}
To assess how well our CMD fitting performed, we generated mock CMDs using known input SFH and extinction values, A$_{V}$ and dA$_{V}$ values. We can then compare the SFH and extinction that are measured in the same way we measure these values for real data. We performed two sets of tests. First, we used realistic SFHs to create mock CMDs. We also ran tests where we used simplistic SFHs with a burst of star formation in a single bin to assess how well the bin and strength of these bursts were recovered.
\subsection{Realistic SFH Tests}\label{sec:realistic_sfh_tests}
\begin{figure*}
\centering
\includegraphics[width=0.95\textwidth]{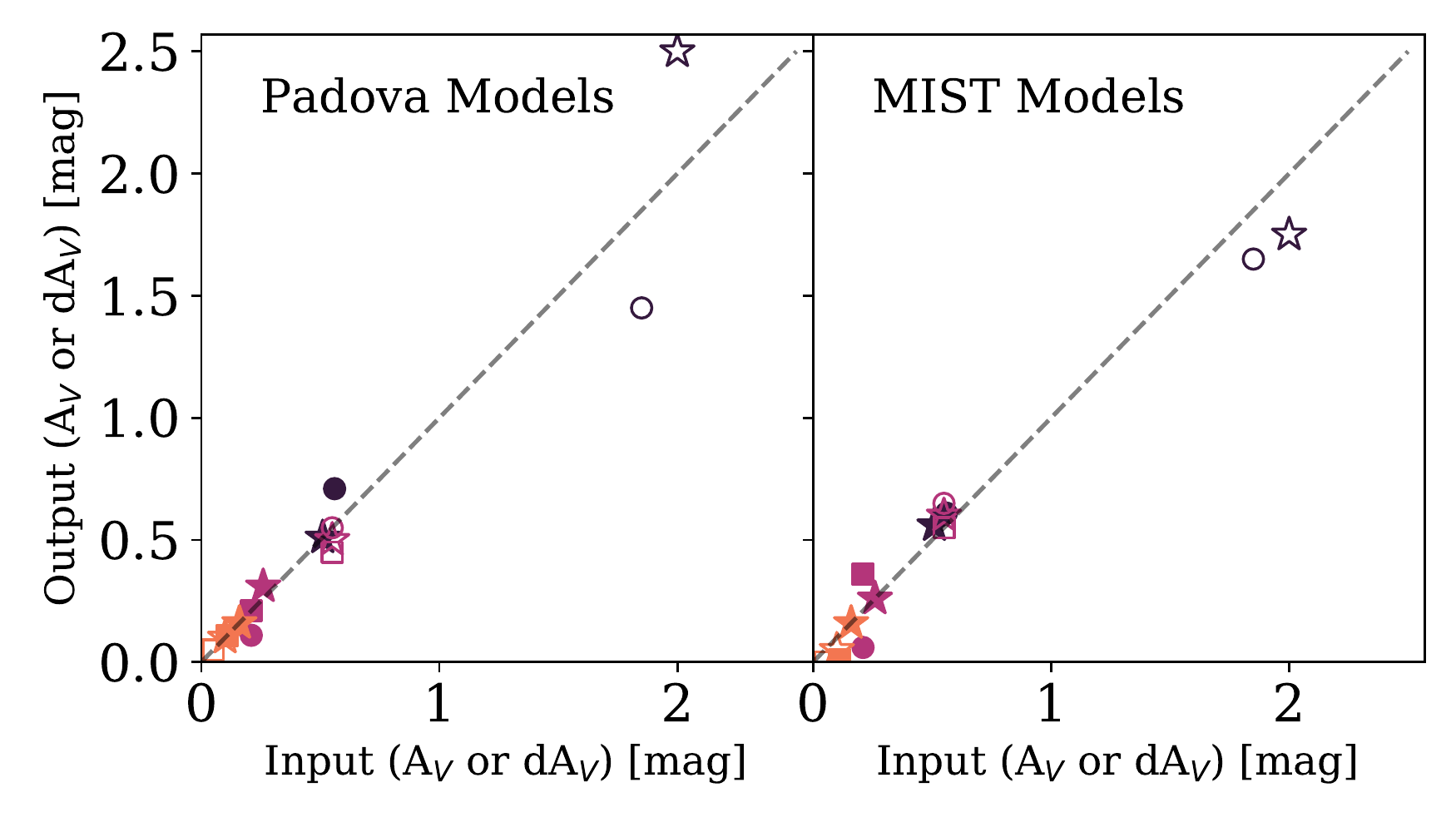}
\caption{Comparison between the input and output A$_{V}$ and dA$_{V}$ values tests where we generated mock CMDs and recovered their extinction and SFH. We created mock CMDs with the input A$_{V}$ and dA$_{V}$ values and then measured these values using the methods described in Section \ref{sec:extinction}. We used the SFH and extinction values for regions from our SFH maps that had varying stellar density and extinction. Filled points indicate A$_{V}$ values and open points indicate dA$_{V}$ values. Dark purple points are high extinction regions, pink points are medium extinction regions, and orange points are low extinction regions. Circles are high stellar density regions, stars are medium stellar density regions, and squares are low stellar density regions.}
\label{fig:av_dav_recovery}
\end{figure*}

\begin{figure*}
\centering
\includegraphics[width=0.99\textwidth]{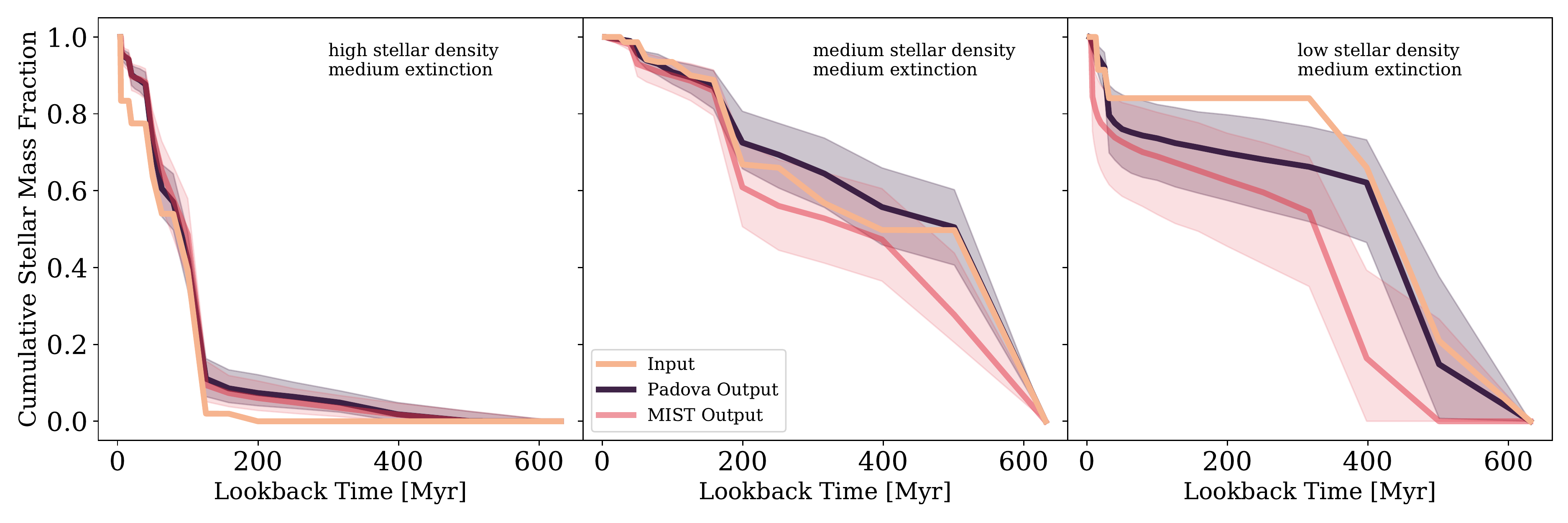}
\caption{Comparison of the cumulative fraction of stellar mass formed over time from tests where we recovered the SFH from mock CMDs generated with realistic SFHs and varying stellar density. The orange line represents the SFH that was used to generate mock CMDs for these regions. The dark purple line represents the SFH recovered from the mock CMD that was fit with the Padova models and the pink line represents the SFH recovered from the mock CMD that was fit with the MIST models. This figure demonstrates the effect of stellar density on our measurements. For the high stellar density region with medium extinction, our errors are fairly small because of the large number of stars in each region on the CMD. The errors are larger, by contrast, in the low stellar density region where the errors are dominated by statistical errors stemming from the low number of stars on the CMD.}
\label{fig:cdf_comparison_dens}
\end{figure*}

\begin{figure*}
\centering
\includegraphics[width=0.99\textwidth]{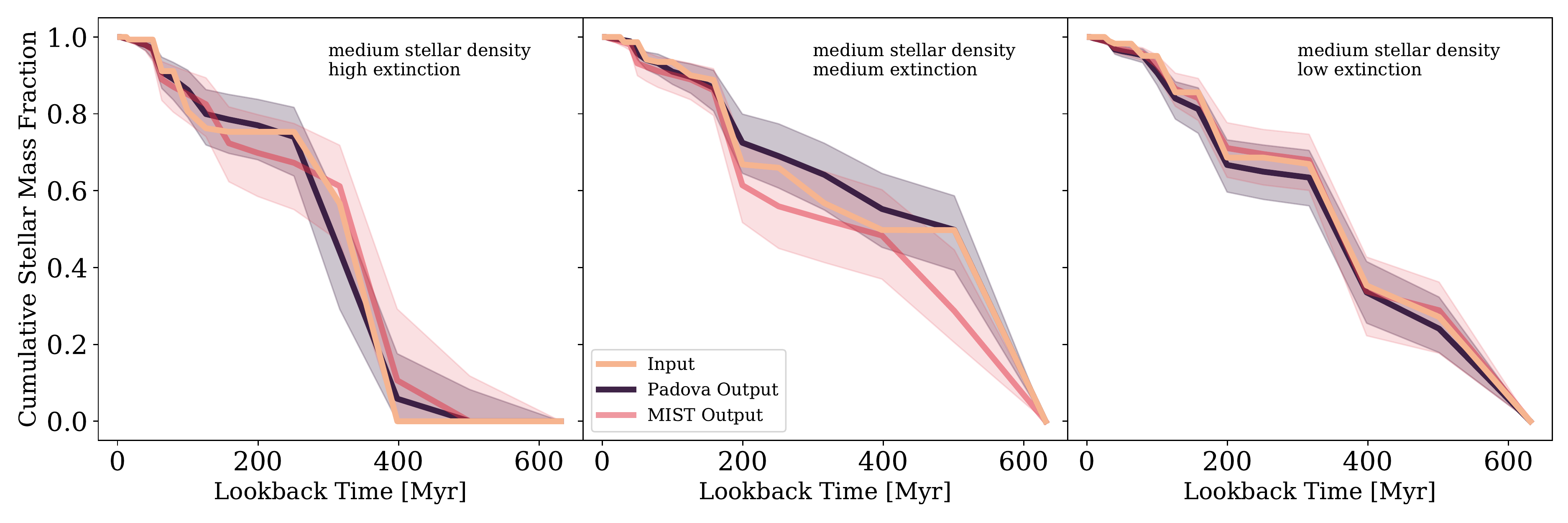}
\caption{Comparison of the cumulative fraction of stellar mass formed over time from tests where we recovered the SFH from mock CMDs generated with realistic SFHs and varying levels of extinction. The orange line represents the SFH that was used to generate mock CMDs for these regions. The dark purple line represents the SFH recovered from the mock CMD that was fit with the Padova models and the pink line represents the SFH recovered from the mock CMD that was fit with the MIST models. This figure demonstrates the effect of extinction on our measurements. The SFH is recovered quite well by both models in all three regions, with smaller errors in the region with low extinction.}
\label{fig:cdf_comparison_ext}
\end{figure*}

For a given test, we chose a region from our SFH maps and used the SFH that was measured for that region as the input to generate a mock CMD. We used that region's best-fit A$_{V}$ and dA$_{V}$ when creating the mock CMD. We then ran this mock CMD through the same pipeline we used to measure our SFHs from the data. We performed this test for regions of varying stellar density and extinction. We chose regions that had combinations of high/medium/low stellar density and high/medium/low extinction. We did not perform these tests for high stellar density/low extinction or low stellar density/high extinction because we did not find any regions in our data that showed this combination of stellar density and extinction, thus these tests would not be representative of our data. We performed these tests with both the Padova and MIST models, using the same model to create the mock CMD and fit for the SFH.
We compared the input versus output A$_{V}$ and dA$_{V}$ values for our test regions. We calculated the root mean square error, which is defined as the square root of the sum of the squares of the differences between the input and output values, scaled by the total number of data points. We found that for the Padova models, $RMSE_{Av}=0.071$ and $RMSE_{dAv}=0.25$. For the MIST models, $RMSE_{Av}=0.093$ and $RMSE_{dAv}=0.13$. In Figure \ref{fig:av_dav_recovery} we plot the input versus output A$_{V}$ and dA$_{V}$ values for our tests with both model sets. Filled points indicate A$_{V}$ values and open points indicate dA$_{V}$ values. Dark purple points are high extinction regions, pink points are medium extinction regions, and orange points are low extinction regions. Circles are high stellar density regions, stars are medium stellar density regions, and squares are low stellar density regions. Both model sets do a good job of recovering A$_{V}$ and dA$_{V}$ values, with the highest extinction regions presenting the most challenge in accurately recovering the dA$_{V}$ value.
We also compare the input and output SFH for these test regions. There are several ways to make this comparison including calculating the median age of the region and comparing the cumulative fraction of stellar mass formed over the time range analyzed in our fits (0-630 Myr ago).
We compared the median age of the SFH used to create the mock CMDs against the median age of the SFH measured with these mock CMDs for both the Padova and MIST models. In this analysis, we use the SFH within the last 630 Myr, the age back to which our fits are reliable. The median age is the age at which 50\% of the total stellar mass formed, and the lower and upper errors come from the ages at which 16 and 84\% of the stellar mass formed. The input and output median ages agree within errors for both model sets, however the Padova models do a slightly better job recovering the median ages for regions with median ages older than about 300 Myr.
We can also compare the cumulative stellar mass fraction formed to compare the input and output SFH over the last 630 Myr. We show these cumulative mass fractions for the input SFH and the output SFH retrieved with the Padova and MIST models in Figures \ref{fig:cdf_comparison_dens} and \ref{fig:cdf_comparison_ext} for regions of varying stellar density and extinction. In Figure \ref{fig:cdf_comparison_dens} all regions have roughly the median value of A$_{V}$ and dA$_{V}$ in our dataset, but vary in their stellar densities. In Figure \ref{fig:cdf_comparison_ext}, all regions have roughly the median stellar density of regions in our dataset, but have varying levels of extinction.
The errorbars on the output cumulative stellar mass fraction in both figures were generated by randomly sampling within the errors on the SFR in each time bin 1000 times. The shaded regions for each model show the 16th and 84th percentile SFH drawn from within the errors.
In summary, we ran tests where we created mock CMDs using a realistic SFH and varying levels of stellar density and extinction. We found that both models recovered the level of extinction quite well, with the MIST models performing slightly better at higher levels of differential extinction. When comparing the median ages of the input SFH versus measured SFH, the Padova models did a slightly better job recovering the median ages towards the older half of our fits. We also compared the cumulative stellar mass fraction for the input SFH and SFHs measured from our mock CMDs. We found that there was not a discernible difference in how well the MIST and Padova models recovered the cumulative stellar mass fraction. We find the largest difference between the input and output SFH for regions with low stellar density, reinforcing the notion that it is better to have a large number of stars on the CMD when recovering the SFH.
\subsection{Burst SFH Tests}\label{sec:burst_tests}
We also ran tests where we used a simplistic SFH with a single burst, or SFR in only one time bin, to generate mock CMDs. We then measured the SFH for these mock CMDs using the same methods we applied to the data, testing our ability to recover the time bin of the burst and the strength of the burst. We ran four sets of burst tests 100 times each. We ran these tests with the Padova models and did not include fitting for A$_{V}$ and dA$_{V}$ because we analyzed our ability to recover these parameters in Section \ref{sec:realistic_sfh_tests}.
We ran four sets of tests where we used a simple burst SFH, which only had SFR in one time bin, to create a mock CMD and test how well the SFR and the bin of the burst were recovered. We ran four sets of burst tests: two early bursts with high/low burst SFR and two late bursts with high/low burst SFR, as described in Table \ref{table:burst_test_info}. The SFR in the bursts were calibrated against the SFRs measured in our observed data, which is how we chose a ``high'' and ``low'' burst strength. Our tests demonstrate that at both early and late times, we recover the time bin of the high SFR bursts within one time bin 100\% of the time. We recover the SFR in the bin within 12\% and 20\% for the early and late bursts, respectively. For the low SFR bursts, we recover the time of the burst within two time bins for the early burst in 100\% of our tests and we recover the time of the burst within two time bins for the late burst in 96\% of our tests. The burst strength recovered in our fits is almost always lower than the injected burst strength, indicating that the stellar mass in the input SFR burst was spread over multiple time bins in the recovered SFH.
\begin{deluxetable*}{cccccccc}
\tabletypesize{\scriptsize}
\tablecaption{Burst SFH Recovery Tests \label{table:burst_test_info}}
\tablehead{
\colhead{Test Name} &
\colhead{Burst Time Bin}  &
\colhead{Burst Time Bin} &
\colhead{Burst SFR} &
\colhead{\% Recovery in}  &
\colhead{\% Recovery Within}  &
\colhead{\% Recovery Within}  &
\colhead{Burst Strength} \\
\colhead{Test Name} &
\colhead{[log(yr)]}  &
\colhead{[Myr]} &
\colhead{[M$_{\odot}$ yr$^{-1}$]} &
\colhead{Exact Time Bin}  &
\colhead{1 Time Bin}  &
\colhead{2 Time Bins}  &
\colhead{Fractional Diff. }
}
\startdata
Early Low & 7.2$-$7.3 & 16$-$20 Myr & 0.002 & 59\% & 91\% & 100\% &$-0.29^{+0.21}_{-0.23}$ \\
Early High & 7.2$-$7.3 & 16$-$20 Myr & 0.03 & 100\% & 100\% & 100\%&$-0.12^{+0.10}_{-0.12}$ \\
Late Low & 8.4$-$8.5 & 251$-$316 Myr & 0.0002 & 40\% & 83\% & 96\% &$-0.91^{+0.03}_{-0.02}$\\
Late High & 8.4$-$8.5 & 251$-$316 Myr & 0.002 & 93\% & 100\% & 100\%& $-0.20^{+0.17}_{-0.22}$\\
\enddata
\tablecomments{Summary of tests where a mock CMD was generated with a burst SFH, which only had SFR in one time bin. Column 1 lists the name of the test. Column 2 lists the time bin of the burst in log(yr) and column 3 lists the start time of the burst in Myr. Column 4 indicates the SFR during the burst. Columns 5$-$7 indicate the fraction of the 100 iterations of a given burst tests during which the burst was recovered in the exact input time bin, within one bin of the input time bin, and within two bins of the input time bin. Column 8 lists the burst strength fractional difference, which is defined as (output burst SFR $-$ input burst SFR)/input burst SFR.}
\end{deluxetable*} 

\end{document}